\documentclass[prb,reprint,showpacs,citeautoscript,floatfix]{revtex4-2}

\pdfoutput=1
\usepackage[charter]{mathdesign}
\usepackage{amsmath}

\usepackage[pdftex]{graphicx}
\usepackage{microtype}
\usepackage{bm}\let\vec\bm

\usepackage[svgnames]{xcolor}
\usepackage[
	colorlinks=True,linkcolor=DarkRed,citecolor=ForestGreen,urlcolor=MediumBlue,
	pdfstartview=FitH,bookmarks=False,pdfpagemode=UseNone
]{hyperref}
\usepackage[english]{babel}
\usepackage[pdftex]{graphicx}
\usepackage{float}
\usepackage{mathrsfs}		
\usepackage{bm}
\usepackage{multirow}
\begin{document}

\title {Propagation of shear stress in strongly interacting metallic Fermi liquids enhances transmission of terahertz radiation}

\author{ D.~Valentinis$^{1,2}$}
\author{J.~Zaanen$^3$}
\author{D.~van~der~Marel$^{1,\ast}$}
 \affiliation{ $^{1}$Department of Quantum Matter Physics, University of Geneva, 24 Quai Ernest-Ansermet, 1211 Geneva 4, Switzerland}
 \affiliation{ $^{2}$Institute for Theoretical Condensed Matter physics, Karlsruhe Institute of Technology, Wolfgang-Gaede Straße 1, 76131 Karlsruhe, Germany}
  \affiliation{ $^{3}$Institute-Lorentz for Theoretical Physics, Leiden University, PO Box 9506, NL-2300 RA Leiden, The Netherlands}
	\email[To whom correspondence should be addressed. E-mail: ]{dirk.vandermarel@unige.ch}

\date{March 29, 2021}
\begin{abstract}
A highlight of Fermi-liquid phenomenology, as explored in neutral $^3$He, is the observation that in the collisionless regime shear stress propagates as if one is dealing with the transverse phonon of a solid. The existence of this $``$transverse zero sound$"$  requires that the quasiparticle mass enhancement exceeds a critical value. Could such a propagating shear stress also exist in strongly correlated electron systems? Despite some noticeable differences with the neutral case in the Galilean continuum, we arrive at the verdict that transverse zero sound should be generic. We present an experimental setup that should be exquisitely sensitive in this regard: the transmission of terahertz radiation through a thin slab of heavy-fermion material will be strongly enhanced at low temperature and accompanied by giant oscillations, which reflect the interference between light itself and the $``$material photon$"$  being the actual manifestation of transverse zero sound in the charged Fermi liquid.
\end{abstract}

\maketitle

\section{Introduction}
The elucidation of the Fermi liquid as a unique state of matter is a highlight of twentieth century physics ~\cite{Nozieres-1999}. It has a precise identity only at strictly zero temperature. At times large compared to $\hbar/(k_BT)$ it can be adiabatically continued to the high-temperature limit and it is therefore indistinguishable from a classical fluid - the $``$collision-full regime$"$ . However, at energies $\hbar \omega >k_B T$ (the $``$collisionless regime$"$) the unique nature of the zero-temperature state can be discerned, being different from either the non-interacting Fermi gas or a thermal fluid. Instead, the closest analogy is with ordered matter characterized by spontaneous symmetry breaking: the Fermi surface takes the role of order parameter, acting however very differently from its $``$bosonic$"$  analogues. A case in point is zero sound. The $F_0^S$ Landau parameter encapsulates the effects of interactions on the compressibility. When this is finite, a propagating collective mode splits off from the Lindhard continuum of particle-hole excitations. The Fermi surface $``$hardens$"$  and zero sound is the coherent s-wave breathing motion of this $``$membrane$"$  corresponding to an oscillation of fermion density in space and time. Historically the neutral Fermi liquid realized in $^3$He has formed the experimental theater to test these general notions. The observation of the sound attenuation maximum at $ \hbar\omega \approx k_B T$ separating the hydrodynamic protection of first (thermal) sound and the $``$Fermi-surface$"$  protected zero sound has been seminal in this regard~\cite{Abel-1966}.

\begin{figure*}[t]
\centering
\includegraphics[width=0.8\textwidth]{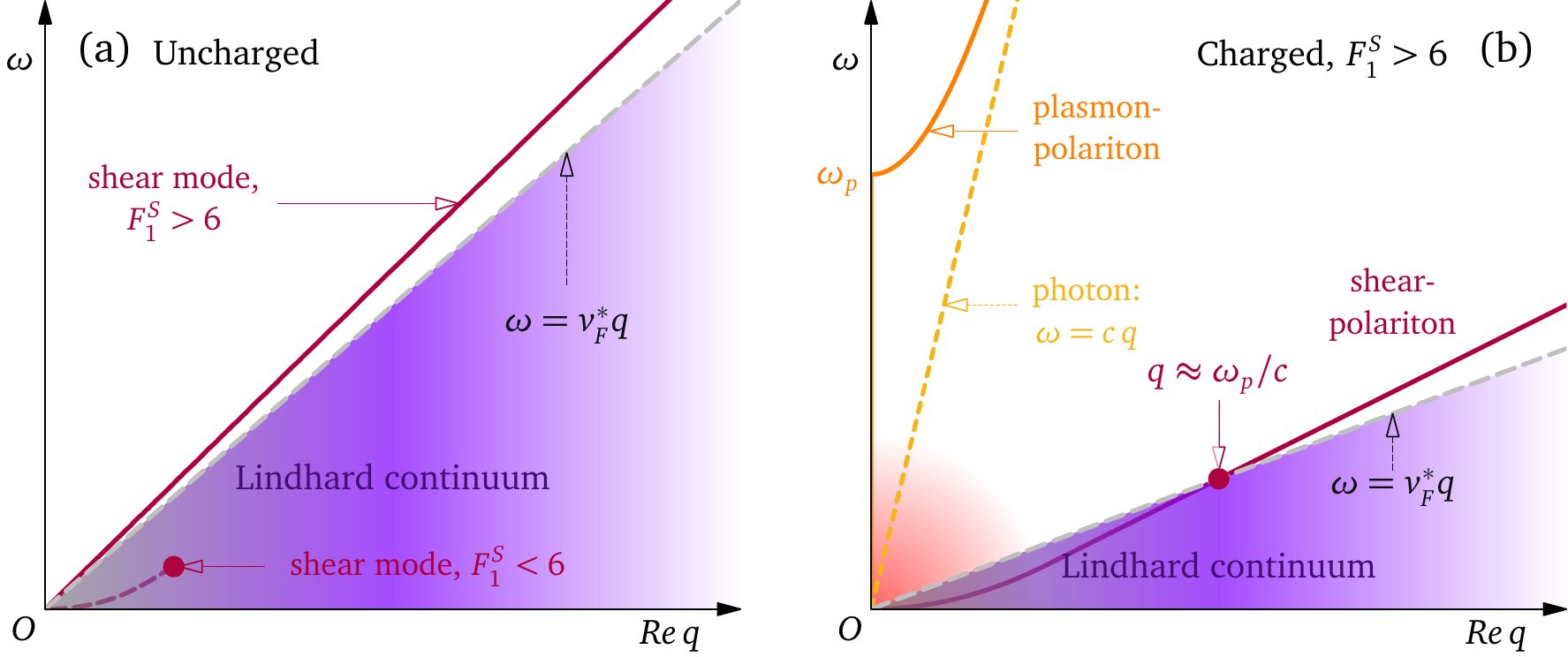}
\caption{
{\bf Transverse collective mode spectrum of neutral and charged Fermi liquids.  } 
\textbf{(a)}  Sketch of the real part of the dispersion relation of transverse sound in a neutral Fermi liquid, corresponding to a pole in the transverse susceptibility. For $F_1^S>6$ (solid line) transverse sound propagates at frequencies above the Lindhard continuum (purple-shaded area); for $F_1^S<6$ transverse sound is Landau-damped (dashed line), and the pole disappears deep inside the continuum at an $F_1^S$-dependent frequency marked by the red dot. \textbf{(b)} Sketch of the real part of the dispersion relation of shear collective modes (polaritons) in a charged Fermi liquid, in the long-wavelength propagating-shear regime with $F_1^S>6$, obtained from the Fermi-liquid transverse dielectric functin in linear response. The interaction of the photon root (dashed golden line) with the transverse sound of the neutral case (panel \textbf{a}) generates the usual plasmon-polariton (solid orange line), which propagates above the plasma frequency $\omega_p$, and the shear-polariton (red solid line), which bears the signatures of Fermi-surface shear rigidity in the charged case. The shear-polariton is repelled by the interaction with the nearby photon root at low frequency (red-shaded area), so that it bends with quadratic dispersion and submerges into the continuum at $q\propto\omega_p/c$. 
\label{fig-omega-q}
}
\end{figure*}

The strongly interacting Fermi liquid has yet another, less well known zero-temperature property. This revolves around the response to external shear stress. Shear rigidity is supposed to be uniquely associated with the breaking of translations, with the ramification that a solid exhibits a propagating shear mode (the transverse acoustic phonon, TA) and a liquid a relaxational response instead. The spatial angular momentum $``$spin-1$"$  Landau parameter regulates the mass enhancement of quasiparticles, $m^\ast/m=1+F_1^S/3$. Abrikosov and Khalatnikov~\cite{Abrikosov-1959,Lea-1973} predicted that for $F_1^S>6$ a new propagating mode forms - see Fig.~\ref{fig-omega-q}a. This mode propagates in fact shear stress ($``$transverse zero sound$"$), entailing a coherent shear deformation of the Fermi sphere~\cite{Lea-1973}. This may be confusing at first sight: the Fermi liquid is a liquid also at zero temperature. The resolution is that the static shear rigidity of a solid is associated with the spin-2 channel involving two space directions ({\it e.g.} Refs.~\onlinecite{Beekman-2017a,Beekman-2017b}). Propagating shear requires only one space direction: it is spin-1 and can therefore be reconciled with a system that is hydrostatically a liquid. Hence, shear-stress propagation in a Fermi liquid is unique in the regard that it is an inherently dynamical phenomenon. As for longitudinal zero sound, temperature dictates the parameter space in which transverse zero sound propagates. The Fermi liquid exhibits a $``$viscoelastic-like$"$  behavior~\cite{Conti-1999}, in the sense that at short times (collisional regime) shear is relaxing while it propagates in the long-time collisionless regime. This prediction was confirmed in the 1970’s by transverse ultrasound measurements in $^3$He~\cite{Roach-1976}.
Electrons in solids also form Fermi liquids. We address here the following question: could it be that propagating shear is ubiquitous in the large variety of heavy Fermi liquid metals~\cite{Stricker-2014}? We will present here the case that it appears to be natural for such collective modes to exist in these systems - see Fig.~\ref{fig-omega-q}b. Propagating shear in solids has been overlooked up to now for the simple reason that it is quite difficult to measure. The difficulty is that one can only exert shear forces on electrons via electromagnetic (EM) fields. To this end, gradients have to be applied and special experimental conditions are required to overcome the kinematic mismatch between the electron and light velocities.
\begin{figure*}[!t]
\centering
\includegraphics[width=0.8\textwidth]{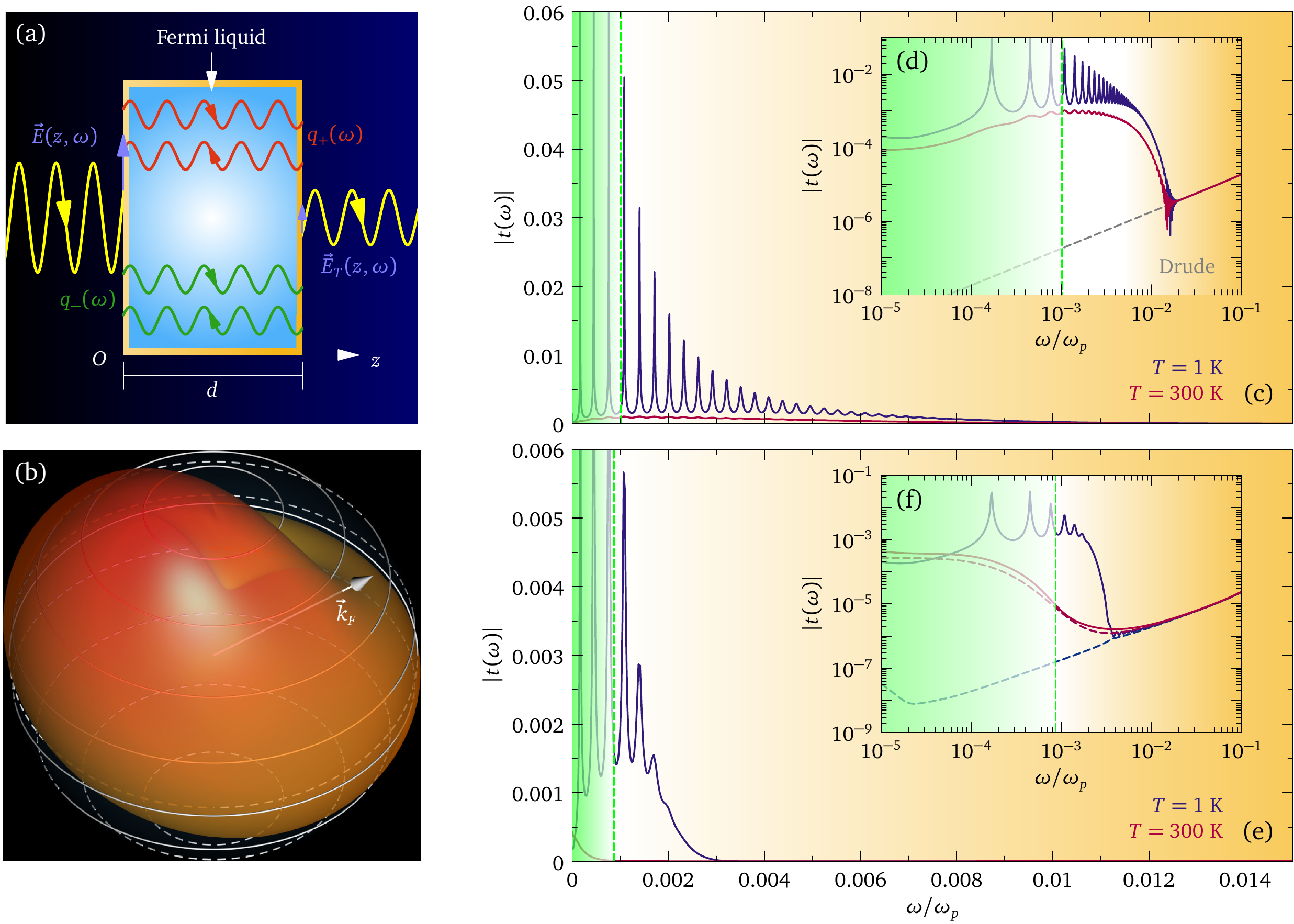}
\caption{
{\bf Simulations of optical transmission experiments at normal incidence on a charged Fermi liquid slab. } 
\textbf{(a)} Schematics of the transmission experiments on a charged Fermi-liquid slab of thickness $d$. The incident electric field $E(z,\omega)$ produces the plasmon-polariton and the shear-polariton in the slab, which correspond to wave vectors $q_+(\omega)$ and $q_-(\omega)$ for each frequency. Each mode is the sum of two counterpropagating rays from Fabry-Perot internal reflections, which lead to the transmitted field $E_T(z,\omega)$. \textbf{(b)} Schematic depiction of the Fermi-surface shear deformation induced by transverse sound. \textbf{(c)} Slab transmission modulus $ \vert t(\omega)\vert = \vert  E_T(z,\omega)/E(z,\omega)\vert $, as a function of frequency $\omega/\omega_p$ for a Fermi liquid in the Galilean continuum at temperatures $T=300$ K (red line) and $T=1$ K (blue line). 
We assume a slab thickness $d=10 c/\omega_p$ ($\sim 0.1 \mu$m for a density $n=10^{23}$ cm$^{-3}$), Landau parameters $F_0^S=1$ and $F_1^S=20$, and the slip length corresponding to specular interface scattering with $hh^{\prime}k_F^2=3.8$ ({\it cf.} sec.\@ \ref{methods}). The shear mode submerges into the continuum below the dashed green line. \textbf{(d)} Same data as in panel \textbf{c} on logarithmic scale. The gray dashed line shows the result for a standard Ohmic conductor. \textbf{(e)} Slab transmission modulus $ \vert t(\omega)\vert$ for a charged Fermi liquid including momentum relaxation due to Umklapp scattering with $\alpha_{U}=0.5$, and acoustic phonons with Debye temperature $ \hbar\omega_{D}/k_{B}=$500 K and electron-phonon coupling constant $\lambda=0.1$. All other parameters are as in panels \textbf{c}-\textbf{d}. \textbf{(f)} Same data as in panel \textbf{e} on logarithmic scale. The gray dashed line is the result for an Ohmic conductor.
\label{fig-t-omega}
}
\end{figure*}

\section{Results}
We will explain an experimental setup where propagating shear should give rise to spectacular, counter-intuitive signals. The configuration is conceptually straightforward: it involves the transmittance of light through a thin metallic layer (see Fig.~\ref{fig-t-omega}a) as a function of radiation frequency and for a fixed layer thickness. Dealing with a weakly interacting Fermi-liquid metal, the incident light is heavily attenuated inside the layer, which becomes essentially nontransparent: the EM field strength is damped exponentially on a length scale set by the skin depth. This behavior continues up to the plasma frequency $\omega_p$. However, upon increasing the quasiparticle mass, a critical point is reached for $m^\ast \geq 3m$:  above this threshold a collective mode of coherent shear deformation of the Fermi sphere propagates inside the material (see Fig.~\ref{fig-t-omega}b), and the transmittance starts to severely oscillate as function of frequency (see Fig.~\ref{fig-t-omega}c-f). In the ideal case of extremely clean samples and very low temperatures, the transmittance can become so large that at the maximum of the oscillation the metallic layer may become quasi-transparent, at a thickness where the weakly interacting metal would block all the incident light (see Fig.~\ref{fig-t-omega}c-d)! Recently Khoo \textit{et al.} studied shear sound of interacting electrons in a different setting, namely in two-dimensional systems~\cite{Khoo-2019} and in narrow strips~\cite{Khoo-2020}.

The plan for the remainder of this paper is as follows. On the basis of the semiclassical Boltzmann kinetic theory applied to the Fermi Liquid (the Landau kinetic equation~\cite{Abrikosov-1959}) we will present here a quantitative phenomenology for the transverse optical response of heavy Fermi-liquid metals in the Galilean continuum. We will firsts re-derive the Abrikosov-Khalatnikov results for the neutral Fermi liquid, which form a useful template for the comparison with the charged Fermi liquid. We will then turn to the charged Fermi liquid, focusing on the implications of shear-stress propagation for the optical properties associated with the transverse response of heavy-mass quasiparticles. We will subsequently compute quantitative predictions for the transmission experiment. We will finish with a discussion of how such computations generalize to heavy-mass Fermi liquids realized in solids. We will suggest an experimental protocol to observe propagating shear modes, concluding with the potential to use this as a new form of high precision $``$Fermi-surface spectroscopy$"$ . The latter may be used to shed a new light on the mysteries of strongly correlated electron systems.

Let us start out reproducing the Abrikosov and Khalatnikov results~\cite{Abrikosov-1959,Lea-1973} for the neutral Fermi liquid. The Landau kinetic equation contains all the information required to analyze the linear response to shear perturbations of given momentum $\vec{q}$ and frequency $\omega$. This information is contained in the transverse susceptibility $\chi_T(\vec{q},\omega)$, while the quasiparticle interactions are parametrized in terms of the momentum-conserving collision time $ \tau_c \approx \hbar E_F/[ (\hbar\omega)^2+(\pi k_B T)^2]$ in the Galilean continuum \footnote{For all calculations in this work we employed the momentum-conserving collision time $\tau_c$ (a single-particle quantity) in accordance with Eq.\@ (24) in sec.\@ \ref{methods}. This is different from the sound-attenuation or optical scattering rates (two-particle quantities), which for a Fermi liquid are $ \propto \hbar E_F/[ (\hbar\omega)^2+(2\pi k_B T)^2]$ \cite{Berthod-2013}.} ({\it cf.} sec.\@ \ref{methods}).
\begin{figure*}[!t]
\centering
\includegraphics[width=0.8\textwidth]{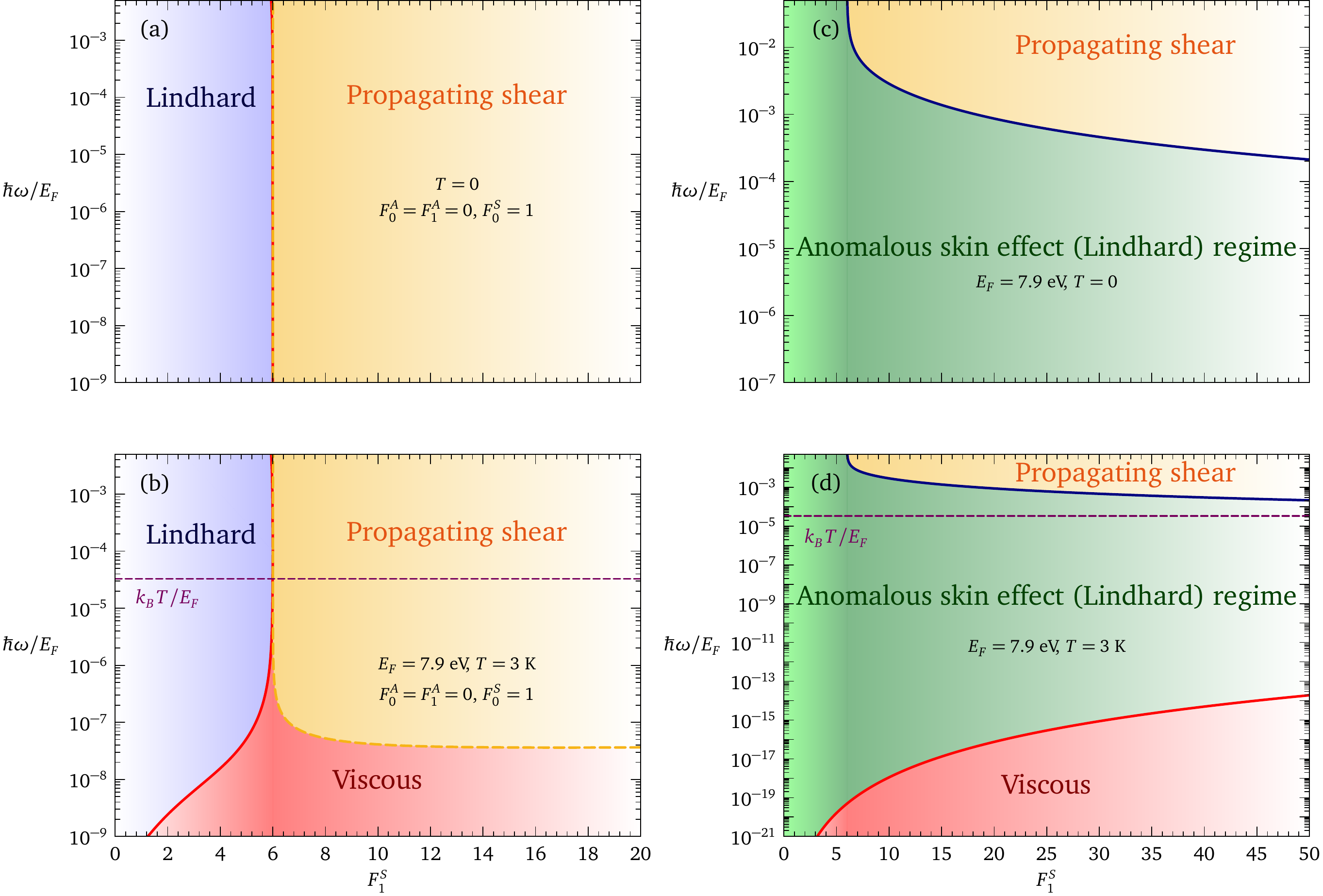}
\caption{
{\bf Diagram of the shear response in neutral and charged Fermi liquids. } 
\textbf{(a)} Shear-response diagram of the zero-temperature neutral Fermi liquid as a function of excitation energy $\hbar\omega /E_{F}$ and $F_1^S$, calculated from the transverse-sound velocity $v_s(\omega)=\omega(q)/q$, and using the Fermi-liquid collision time $\tau_c$. Notice that the zero-temperature phase diagram is valid at any electron density $n$, since at $T=0$ the collision time $\tau_c$ becomes a universal function of $\hbar\omega /E_F$. The threshold $F_1^S=6$ separates the Lindhard regime, where the Lindhard continuum dominates the response and the transverse susceptibility has no pole, and the propagating shear regime, where transverse sound propagates. \textbf{(b)} Neutral shear-response diagram at $T=3$ K and $E_F=7.6$ eV (electron density $n=10^{23}$ cm$^{-3}$). For $F_1^S<6$, the red solid line marks the crossover between the viscous regime of hydrodynamic damped transverse sound and the Lindhard regime. For $F_1^S>6$, the condition $\omega=v_F^\ast$Re$q$ (dashed golden line) identifies the crossover between viscous and propagating shear regimes. \textbf{(c)} Long-wavelength shear-response diagram of the charged Fermi liquid at $T=0$ and $E_F=7.6$ eV, calculated from the dispersion relation of the shear-polariton and the Fermi-liquid optical conductivity in Kubo formalism. The shear-polariton emerges from the continuum at the solid blue line. Deep inside the continuum, the response is dominated by the Lindhard continuum and yields the phenomenology of anomalous skin effect. \textbf{(d)} Charged shear-response diagram at $T=3$ K and $E_F=7.6$ eV. The red line shows the condition $\delta_0=l_c$, with $\delta_0$ the EM skin depth in anomalous regime and $l_c$ the collisional mean free path, which indicates the crossover between the anomalous skin effect and viscous regimes, with the latter implying hydrodynamic response.
\label{fig-omega-F}
}
\end{figure*}
\section{Discussion}
Transverse zero sound appears as a pole of $\chi_T(\vec{q},\omega)$ above the critical $F_1^S$, representing the oscillating shear deformation of the Fermi surface (see Fig.~\ref{fig-t-omega}b). This reproduces the main Abrikosov-Khalatnikov result as indicated in Fig.~\ref{fig-omega-q}a: the shear mode appears as an anti-bound state formed from the Lindhard continuum. In the collisionless limit $\tau_{c} \rightarrow +\infty$ and at large $F_1^S$, this mode has a velocity $v_s=v_F\sqrt{(1+F_1^S/3)/5}$~\cite{Valentinis-2020}, which is enhanced with respect to the velocity $v_F^\ast$ of Fermi-surface quasiparticles. The quasiparticle Fermi velocity is renormalized by interactions according to $v^\ast_F=\hbar k_F/m^\ast = v_F/(1+F_1^S/3)$ where $v_F=\hbar k_F/m$ is the bare Fermi velocity and $k_{F}=(3\pi^2 n)^{1/3}$ is the Fermi wave vector in a three-dimensional Fermi liquid, with $n$ electron density. Therefore, for small $q$ and $F_1^S>6$ transverse sound is protected against the Landau damping associated with entering the Lindhard electron-hole continuum~\cite{Abrikosov-1959}. We notice that for $F_1^S<6$ a remnant of the shear mode still exists at small momenta, but it is critically damped due to the decay in the Lindhard continuum (dashed line in Fig.~\ref{fig-omega-q}a), and it ceases to exist altogether deep inside the continuum (dot in Fig.~\ref{fig-omega-q}a)~\cite{Lea-1973}.

For future purposes it is convenient to sketch a $``$phase diagram$"$  of the shear response in the neutral Fermi liquid as a function of $\omega$  and $F_1^S$, both for zero- (\textbf{a}) and finite temperature (\textbf{b}): Fig.~\ref{fig-omega-F}. This will later serve as a basis to understand the transverse response of quasiparticles endowed with electric charge, where the latter allows for the interaction of Fermi-surface shear modes with photons.

In essence, the response diagram of Fig.~\ref{fig-omega-F}a-b outlines the conditions on $\omega$  and $F_1^S$ for transverse sound to develop in the neutral system, as testified by the emergence of a pole in the susceptibility $\chi_T(\vec{q},\omega)$ ({\it cf.} sec.\@ \ref{methods}) in the presence of the Fermi-liquid collision time $\tau_{c}$~\cite{Lea-1973,Roach-1976,vanderMarel-2011}. An analysis of the mode dispersion relation $\omega(q)$ for $F_1^S>6$ confirms that the real (\textit{i.e.} reactive) part of $\omega(q)$ prevails over its imaginary (i.e. dissipative) part for Re$\omega(q)\ge v^\ast_F q$: indeed transverse sound propagates nearly undamped outside of the Lindhard continuum, consistently with the sketch in Fig.~\ref{fig-omega-q}a. Such propagation occurs in the regime $\omega\tau_{c}\gg 1$, and in the large-$F_1^S$ collisionless limit the mode velocity reaches $v_s^\infty$.

It is instructive to translate the qualitative criterion Re$\omega(q)=v^\ast_F q$, which characterizes the crossover from overdamped to propagating shear at the edge of the continuum, into a condition Re$\omega(q)\ge v^\ast_F q$ on the mode velocity $v_s(\omega,F_1^S)=\omega(q)/q$ as commonly defined in the $^3$He literature~\cite{Lea-1973,Roach-1976}. We determine such condition numerically, as marked by the dashed orange line in Fig.~\ref{fig-omega-F}a-b.
Conversely, for $F_1^S<6$ the pole in $\chi_T(\vec{q},\omega)$ is heavily Landau-damped at low frequency, in the hydrodynamical/collisional regime $\omega\tau_{c}\ll1$, and it disappears at an $F_1^S$-dependent frequency evaluated numerically and marked by the red solid line in Fig.~\ref{fig-omega-F}a-b.

At zero temperature (Fig.~\ref{fig-omega-F}a) the physical meaning of the above analysis is very simple: above the critical $F_1^S$ the response is governed by the propagating shear mode, and otherwise one recovers the incoherent response of the Fermi gas, devoid of collective modes. In the finite-temperature case (Fig.~\ref{fig-omega-F}b) the crossover from the high-frequency zero sound to the low-frequency thermal regime occurs at a frequency $ \hbar \omega \ll k_{B}T$ , as known in the $^3$He community: such frequency is so low because we need many collisions during an excitation cycle, {\it i.e.} $ \omega  \tau_{c}\ll 1$, to establish local equilibrium among quasiparticles and enable hydrodynamic behaviour. In thermal regime there is no longer a critical change associated with $F_1^S$: we are dealing with the Galilean-invariant continuum, which is always governed by Navier-Stokes hydrodynamics characterized by the viscosity of the finite-temperature Fermi liquid $\nu(0)=v_F^2\tau_c$, as follows from the kinetic theory~\cite{Abrikosov-1959,FZVM-2014} ({\it cf.} sec.\@ \ref{methods}).
With regard to hydrodynamics, it will become clear that it is largely irrelevant for the electron systems: extremely clean systems such as graphene are required to observe hydrodynamic flow~\cite{Crossno-2016,Bandurin-2016,Kumar-2017,Gooth-2018,Sulpizio-2019,Ella-2019}, while the latter is actually very difficult to detect by radiative means.

Let us now turn to the less well-charted charged Fermi liquid - to the best of our knowledge the transverse response for large $F_1^S$ was only addressed in the book by Nozieres and Pines~\cite{Nozieres-1999} in the form of an exercise. A first ramification of the Coulomb interactions is a renormalization of the Landau parameters~\cite{Silin-1958a,Silin-1958b}. More importantly, one has now to address the linear response to EM sources, which can be handled by time-dependent mean-field (RPA) using the Kubo formalism ({\it cf.} sec.\@ \ref{methods}). The way this works in the longitudinal channel is well known. The zero (longitudinal) sound mode gets $``$dressed$"$  by the Coulomb interaction, with the effect that it is promoted to a plasmon with dispersion large $\sim\sqrt{\omega_p^2+v_F^2q^2}$ where $\omega_p=\sqrt{n e^2/\epsilon_0 m}$ is the electron plasma frequency.

However, light is transversally polarized. This means that, in the presence of a propagating transverse mode in the material, the latter blends with light through radiation-matter interaction, which can be rationalized in terms of simple linear mode coupling~\cite{Forcella-2014}. As for the plasmon, this is a universal affair and one may as well first consider a Wigner crystal formed by electrical charges~\cite{Beekman-2017a}. The shear modes of such ensemble are the electronic analogues of transverse acoustic phonons in ionic lattices, and such $``$phonons$"$  couple to photons possessing a much larger velocity (of light). A first product of this coupling is the usual $``$light-like$"$  transversally polarized plasmon-polariton, which is universal: indeed, at $q=0$ both transverse and longitudinal $``$plasmons$"$  precisely match. However, its partner is the phonon-like $``$shear-polariton$"$: this mode continues to be massless, but it acquires a quadratic dispersion $\omega\sim q^2$ at small momenta~\cite{Beekman-2017a,Beekman-2017b} (a similar mechanism occurs in a particular viscoelastic-like holographic strange metal, see reference \onlinecite{Baggioli-2020}.) The shear mode of the heavy-mass Fermi liquid shares an analogous fate, see Fig.~\ref{fig-omega-q}b. The only qualitative difference with the neutral case of Fig.~\ref{fig-omega-q}a is that, for very small momenta being less than $q \approx\omega_p/c$ ({\it cf.} sec.\@ \ref{methods}), the shear mode dips below the upper bound of the Lindhard continuum at $\omega\approx v_F q$, thus falling prey to Landau damping. Actually, according to the kinetic theory, Landau damping further modifies the mode dispersion, which becomes  $\omega\propto q^4$ at vanishing frequencies~\cite{Valentinis-2020}.

Once again, the only way to exert forces at finite frequency on the charged Fermi liquid is by electromagnetic radiation. This in turn implies restrictions on which intrinsic properties of this system can actually be measured: a primary consequence is that the response to the EM field is inherently a boundary phenomenon. The reason is that an oscillating EM field is screened by the metal, with the ramification that the transverse response corresponding to the screening currents decreases exponentially away from the boundary, the classical skin effect.
In an Ohmic conductor the skin depth is $\delta_s=\sqrt{2\rho/\mu\omega}$ where $\rho$ and $\mu$  are the DC resistivity and permeability of the conductor, respectively. This in turn assumes the Drude model, departing from a free Fermi gas, characterized by an overall scattering time $\tau_K$ due to momentum relaxation. In the Galilean continuum this is absent, and one has to reconsider the skin effect mechanism.
The answer is well known at low temperature and small $F_1^S$ in the form of the anomalous skin effect, elucidated a long time ago by Reuter and Sondheimer~\cite{Reuter-1948}. Upon lowering temperature, the mean free path of the electrons $l_{mf}=v_F \tau$ (either $\tau_c$, $\tau_K$ or both) exceeds the length over which the transverse electric field penetrates the metal. The screening is now entirely due to the Lindhard continuum, with the outcome that the skin depth saturates as a function of temperature at a value 
$\delta_0=\{(4c^2 v_F)/(3\pi\omega\omega_p^2)\}^{1/3}$.

This implies in turn that in the Galilean continuum the finite-temperature hydrodynamical fluid ({\it e.g.} Fig.~\ref{fig-omega-F}b) is actually hidden from observation in a large frequency regime! Hydrodynamics sets in at a scale larger than the collision length $l_c= v_F^\ast \tau_c$, but we just learned that the screening is complete at the length scale $\delta_0<l_c$. In this regime the hydrodynamical fluid responds therefore like a free Fermi gas. Upon raising temperature $l_c$ decreases, and when $\delta_0\approx l_c$ a crossover occurs to a regime where hydrodynamical currents take over the screening ({\it cf.} sec.\@ \ref{methods}). In an earlier paper we analyzed the ramifications of such screening~\cite{FZVM-2014}. These observations may be of relevance in the context of graphene-like systems, but we leave a detailed analysis to a future publication.

As for the neutral case, we can sketch an electromagnetic response diagram: Fig.~\ref{fig-omega-F}c-d. In the presence of electric charge, the role of $\chi_T(\vec{q},\omega)$ is taken by the transverse dielectric function $\epsilon_T(\vec{q},\omega)$ , with the two quantities linked by the Kubo formula ({\it cf.} sec.\@ \ref{methods}). The fact that the Fermi liquid is interrogated by photons, with dispersion relation \textit{$\omega$ =c q} (see Fig.~\ref{fig-omega-q}b), fixes momentum so that the only independent parameters are $\omega$  and $F_1^S$. Polaritons are self-consistent solutions of Maxwell’s equations in the Fermi liquid, {\it i.e.} $q^2c^2/\omega^2=\epsilon_T(\vec{q},\omega)$: in other words, we look at the photon self-energy to deduce the response of the matter.

In the following, we describe the EM phenomenology through the analytical solutions for polaritons obtained deep in the low- and high-frequency regimes ({\it cf.} sec.\@ \ref{methods}), which are sufficient to understand the essential physics. We leave the full numerical computation of polaritons near crossovers between different regimes for a subsequent work, but here we provide robust qualitative estimates on where such crossovers occur.
In the high-frequency, long-wavelength regime $\omega\gg v_F^\ast$Re$q$, at leading order we find two polariton branches ({\it cf.} sec.\@ \ref{methods}): one is the usual plasmon-polariton root, which propagates above the plasma frequency. The other, lower-frequency root is the $``$shear-polariton$"$ , a genuine product of Fermi-surface shear interacting with light. As such, the shear-polariton is a collective mode rooted in time-dependent mean field, being in this regard in the same category as the plasmon or even a TA phonon. Via continuity, the shear-polariton is associated with a coherent shear current, which takes over the screening from the Lindhard excitations as soon as it springs into existence.

Fig.~\ref{fig-omega-F}c-d shows that, compared to the neutral case, charged propagating shear takes over above a critical, finite frequency. This expresses the fact that at large wavelengths the shear mode becomes overdamped because of its dispersion bending from linear to quadratic due to EM forces at $q\le\omega_p/c$ (see Fig.~\ref{fig-omega-q}b), which causes the mode to dive into the Lindhard continuum~\cite{Nozieres-1999}. The qualitative criterion for the shear-polariton to emerge from the continuum is that its dispersion obeys $\omega\gg v_F^\ast$Re$q$. We estimate the latter condition analytically ({\it cf.} sec.\@ \ref{methods}), as shown by the blue line in Fig.~\ref{fig-omega-F}c-d. Notice that the shear-polariton never propagates for $F_1^S<6$: this is because light needs to couple to propagating transverse sound in the material to generate the propagating shear-polariton, and transverse sound propagation necessitates $F_1^S>6$ as previously explained.

On the other hand, in the low-frequency, short-wavelength regime $\omega\ll v_F^\ast$Re$q$ and at finite $\tau_{c}$, the leading-order expansion of the Fermi-liquid dielectric function yields anomalous skin effect ({\it cf.} sec.\@ \ref{methods}) due to the incoherent response of the continuum to EM fields.
A final difference with the neutral case is that the apparent onset of the hydrodynamical $``$viscous$"$  regime is at a much lower frequency than in the neutral case. But as we explained, this is in a way a deception: although in principle the bulk may well already be in the thermal regime, the boundary layer where screening currents are running is still too thin for the fluid to reach local thermal equilibrium, given that the collision length is large compared to the skin depth, and the ensuing Fermi-gas response (anomalous skin effect) is insensitive to the difference between the thermal- and zero-temperature liquid. We qualitatively estimate the crossover between Lindhard and hydrodynamical regimes by solving $\delta_0=l_c$ numerically with the Fermi-liquid collision time $\tau_{c}$: this gives the red curve in Fig.~\ref{fig-omega-F}d.

Having established where to look for the propagating shear in the Fermi-liquid metal, the next question is how to find it. In this regard we wish to propose an experiment yielding a unique, qualitative signal that will unambiguously confirm or falsify the existence of the shear mode. This revolves around shining light through a thin, micrometer-sized slab of metal.
In a normal metal, the field strength of the radiation decays on the length scale set by the (anomalous) skin depth, which is small compared to the width of the slab. In these conditions, the slab is opaque. However, the propagating shear mode is in essence a photon $``$made from matter$"$, being only different in the regard that its velocity is much smaller than the light velocity. The coupled light-matter system is therefore characterized by two propagating modes: the light-like plasmon- and matter-like shear-polariton, characterized by wave vectors 
$q_+(\omega)$ and $q_-(\omega)$ -see Fig.~\ref{fig-t-omega}a. 
These modes propagate with different velocities, while both are in part reflected back at each interface. Given the difference in the real part of their momenta, their phases evolve differently along the path through the material. Accordingly -- pending the frequency of the incoming light -- the two polaritons interfere constructively or destructively at the slab boundaries, giving rise to a transmitted field that exhibits oscillations as function of frequency (see Fig.~\ref{fig-t-omega}b). Given the propagating nature of both polaritons, one already anticipates that these oscillations in the transmittance of the slab may be very large, as confirmed by our computations: Fig.~\ref{fig-t-omega}c-d.

We addressed a similar set up some time ago, dealing instead with a Navier-Stokes liquid characterized by its relaxational shear mode~\cite{FZVM-2014}. This computation is trivially adapted to the present propagating shear -- it is just an exercise involving the transmission and reflection of two linearly coupled evanescent waves in the slab ({\it cf.} sec.\@ \ref{methods}). In the hydrodynamical case the shear mode is overdamped, with a complex momentum dominated by its imaginary part while it has still a finite real ($``$propagating$"$ ) part. We found out that this sufficed to form similar oscillations as function of frequency, being however very small while the transmission as a whole is strongly attenuated~\cite{FZVM-2014}. Dealing with the propagating shear mode, its imaginary momentum is small compared to its real one, which entails that the slab becomes quite transparent and characterized by large oscillations in the transmittance as a function of frequency.

Using the results of the kinetic theory as an input, we can realistically simulate the transmission of slabs containing Galilean-invariant Fermi liquids ({\it cf.} sec.\@ \ref{methods}). For the results in Fig.~\ref{fig-t-omega} we assumed a slab thickness $d=10c/\omega_p$ ($\cong$ 0.1 $ \mu$m for a density $n=10^{23}$ cm$^{-3}$), Landau parameters $F_0^S=1$, and $F_1^S=20$, and the slip length corresponding to specular interface scattering with $hh^{\prime}k_F^2=3.8$ allowing for a maximal transmission of $ \vert t(\omega)\vert\approx $0.06 in the frequency range between $10^{-3}\omega_p$ and 0.1$\omega_p$. As we just discussed, at small frequencies up to the green dashed line the shear mode is overdamped, while spectacular transmittance oscillations set in upon entering the propagation regime (see blue line in Fig.~\ref{fig-t-omega}d). One infers that the overall transmission is much larger than the standard result for an ohmic conductor (dashed gray line in Fig.~\ref{fig-t-omega}d). Up to this point we ignored the lifetime of the shear mode. This is accounted for by the kinetic theory, parametrized by the frequency dependence of the collision time, $\tau_c\propto \omega^{-2}$. This attenuates the shear mode at high frequencies, with the consequence that the enhanced transmission decreases rapidly as illustrated in the log-log plot (Fig.~\ref{fig-t-omega}). Raising temperature has a very similar effect since $\tau_c(\omega=0,T)\propto T^{-2}$.  (red line in Fig.~\ref{fig-t-omega}d).

The conclusion is that, in order to detect the propagating shear oscillations in a heavy-mass Fermi liquid, temperature should be small compared to $E_F$, while small frequencies (of order of  $10^{-3}$ of the plasma frequency) are required, just above the point where the propagating shear escapes the long-wavelength Landau damping sketched in Fig.~\ref{fig-omega-q}b.

As illustrated in the above, the Landau kinetic theory has remarkable powers in its capacity to deliver a quantitative phenomenology. However, it requires Galilean invariance, which is badly broken in the experimentally available charged Fermi-liquids: the metallic state of electrons formed in solids. Such breaking of Galilean invariance has the fundamental ramification that total momentum is no longer conserved. Considering the longitudinal response of the Fermi-liquid, total momentum conservation entails that the sound mode (or the plasmon) is infinitely long-lived both in the zero- and finite-temperature cases in the long-wavelength limit. The same principle applies to transverse zero sound, although it needs in addition the sharp zero-temperature Fermi surface.

The role of momentum (non-)conservation is thereby identical regardless of whether one is dealing with the longitudinal or transverse electromagnetic response, and we know precisely how this works for the former. In the Fermi liquid, the large Umklapp momentum scattering associated with a perfectly periodic lattice is irrelevant at long wavelength, and what remains is elastic scattering against impurities. Furthermore, at finite temperature (and energy) momentum relaxation also acts through electron-electron interactions and electron-phonon coupling. Good metals are $``$nearly hydrodynamical$"$ , in the sense that the width of the Drude peak, which represents the momentum relaxation rate, is small. Such scale governs the decay of transverse sound as well. In a Fermi liquid, the main consequence of a finite-momentum lifetime is that the hydrodynamical flow of the continuum is in any but the most perfect solids (like graphene) immediately destroyed and replaced by the Drude (Ohmic) regime, where momentum is damped in the lattice by the individual quasiparticles~\cite{Gurzhi-1968}. Consequently, the hydrodynamic regime in Fig.~\ref{fig-omega-F}d is replaced by the Ohmic skin effect.

However, the propagating shear is only affected $``$perturbatively$"$ : it will decay faster compared to the continuum case due to momentum relaxation. To offer some intuition regarding these numbers, we show in Fig.~\ref{fig-t-omega}e-f a realistic example ({\it cf.} sec.\@ \ref{methods}). Compared to Fig.~\ref{fig-t-omega}c-d we introduce two additional scattering channels: the temperature- and frequency-dependent momentum relaxation time $\tau_{K}(T,\omega)$ which parametrizes the $T^2$ contribution rooted in a realistic $``$Umklapp efficiency$"$ $\alpha_{U}=0.5$~\cite{FZVM-2014,Valentinis-2020}, and the coupling to phonons, which is particularly efficient in relaxing momentum. Transverse zero sound will couple to transverse lattice vibrations, and we consider such acoustic phonons with a Debye temperature $ \hbar\Omega_D/k_{B}=$ 500 K and an electron-phonon coupling constant $\lambda=0.1$. The outcome is that phonons and Umklapp effectively conceal any signal of shear propagation at high temperatures, but at low temperatures the oscillations are still clearly discernible. However, even for a relatively weak electron-phonon coupling, the phonons represent an efficient decay channel for transverse zero sound: as the figure suggests, one better looks for transverse shear at frequencies below the typical phonon frequency scale in a given material.

In addition to slab transmission, surface impedance measurements provide in principle simpler access to the EM skin depth which discerns the different regimes in Fig.~\ref{fig-omega-F}d~\cite{FZVM-2014,Valentinis-2020}. However, explicit computation~\cite{Impedance-unpubl-2020} shows that the differences between the signals in propagating-shear, Ohmic and anomalous skin effect regimes are minimal, once momentum relaxation is taken into account. Hence surface impedance experiments on bulk heavy-mass Fermi-liquids are only subsidiary with respect to thin-film optical spectroscopy in identifying Fermi-surface propagating shear.

A real complication is that invariably strong periodic background potentials are required to endow the Fermi surface of electron systems with a large quasiparticle mass. In turn, these strongly renormalized Fermi liquids are rooted in a strongly interacting microscopic physics, which is typically described by Kondo lattice models, by the proximity to quantum phase transitions, and so forth~\cite{Coleman-2015}. As a matter of fact, it is not at all understood why a heavy Fermi liquid emerges in the deep infrared, but this is the domain where the experimental study of transverse zero sound may have considerable potential as a spectroscopic technique, due to its exquisite sensitivity to the characteristics of the Fermi surface.

There is leeway for theoretical guidance in the form of the tradition on determining the renormalized band structure governing the low-temperature Fermi-surface topology. This is typically a rather complex affair for real solids, with a variety of sheets characterized by different effective masses~\cite{Kotliar-2006}. In such case, the shape of the Fermi surface, as well as the quasiparticle mass and transverse sound, are associated with projections based on the point-group symmetries of the crystal. This implies that, for instance, the velocity and the damping of transverse sound become anisotropic, but this can be dealt with in principle, as demonstrated in the theory of elasticity and quite recently also in hydrodynamics~\cite{Bradlyn-2012,Cook-2019}.

Can transverse zero sound be measured in the laboratory, to be exploited as a spectroscopic tool to study the mysteries of electron systems with heavy quasiparticle mass? From the above we have established the required conditions. First, one needs good metals characterized by a small resistivity at low temperatures, to assure that transverse sound does not decay too rapidly through momentum relaxation. To avoid complications associated with extreme anisotropy, the ideal candidate materials for transverse-shear optical spectroscopy should be preferably rather isotropic and three-dimensional, with a band structure as simple as possible. 
Most importantly, they should be characterized as Fermi liquids with large quasiparticle mass enhancement: since a large $F_1^S$ protects transverse sound from Landau damping, the higher the mass the better. Examples of candidate materials fulfilling these criteria are $\alpha$-cerium~\cite{vanderEb-2001,Haule-2005}, MnSi~\cite{Mena-2003}, and UPd\textsubscript{2}Al\textsubscript{3}~\cite{Scheffler-2005}. These should be tailored into submicron-width thin layers and then cooled to low temperature, while subsequently the transmission should be measured as function of frequency in the terahertz range: the sweet-spot frequency window to look for lies in between the low-frequency regime, where the traditional skin effect takes over, and the regime where phonons damp transverse sound. It is a beneficial circumstance that the technology to realize these conditions in the laboratory has just become available~\cite{Prochaska-2020}.

\acknowledgements
We acknowledge stimulating discussions with A. Morpurgo, C. Berthod, J. Schmalian, A. B. Kuzmenko, G. A. Inkof, M. Baggioli and F. Pientka. 
This work was supported by the Swiss National Science Foundation through grants no. 200020$-$ 179157 (D.v.d.M.), and the Early Postdoc Mobility grant P2GEP2 181450 (D.V.). \\
\noindent{\bf Author contributions} \\
DV, JZ and DvdM conceived the project and discussed the model, DV performed calculations, DV, JZ and DvdM wrote the paper. \\
\noindent{\bf Competing interests} \\
None. \\
\noindent{\bf Materials \& Correspondence} \\
Not applicable. 


\section{Methods}\label{methods}

In this section we summarize the main features of the model used to obtain all results discussed in the main text. Additional details of the formalism are described in Ref. \onlinecite{Valentinis-2020}. 	\\ \\
\noindent
\subsection{Landau kinetic equation}

We start from the Landau kinetic equation, which determines the first-order deviation (or ``displacement") $\epsilon_{\vec{k}}(\vec{q},\omega)$ of the quasiparticle distribution function, with momentum $\vec{k}$ on the Fermi surface, with respect to global thermodynamic equilibrium, driven by external perturbations of momentum $\vec{q}$ and frequency $\omega$. Perturbation sources include quasiparticle interactions, collisions, momentum-relaxing scattering and external driving forces like electromagnetic fields. In general, the kinetic equation reads~\cite{Nozieres-1999,Abrikosov-1959,Valentinis-2020}
\begin{multline}\label{eq:kin_k_pot}
\left(\vec{q} \cdot \vec{v}_{\vec{k},\sigma}-\omega \right) \epsilon_{\vec{k}}(\vec{q},\omega)  +\vec{q} \cdot \vec{v}_{\vec{k},\sigma} \delta \left[ E\left(\vec{k},\sigma \right)- \mu \right] \\ \cdot \left\{ \sum_{\vec{k}',\sigma'} \textsl{f}_{\vec{k},\sigma, \vec{k}', \sigma'} \epsilon_{\vec{k}'} (\vec{q},\omega) \delta\left[ E_n(\vec{k}',\sigma')-\mu \right]  \right\}  \\ ={\mathscr{I}}_{coll}(\vec{q},\omega)+{\mathscr{I}}_{r}(\vec{q},\omega) \\ -\frac{e}{m v_F^\ast} \left(\vec{q} \cdot \vec{v}_{\vec{k},\sigma}\right) \delta \left[ E\left(\vec{k},\sigma \right)- \mu \right] \vec{k} \cdot \vec{A}(\vec{q},\omega),
\end{multline}
where $\vec{v}_{\vec{k},\sigma}$ is the quasiparticle velocity for momentum $\vec{k}$ and spin $\sigma=\left\{\left|\uparrow\right\rangle, \left|\downarrow\right\rangle\right\}$, $\textsl{f}_{\vec{k},\sigma, \vec{k}', \sigma'}$ encode short-ranged interactions between quasiparticles, ${\mathscr{I}}_{coll}(\vec{q},\omega)$ is the momentum-conserving collision integral, ${\mathscr{I}}_{r}(\vec{q},\omega)$ is the integral for momentum-relaxing scattering, $\vec{A}(\vec{q},\omega)$ is the applied vector potential, and the delta functions $\delta\left[ E_n(\vec{k},\sigma)-\mu \right]$ select quasiparticle states at the chemical potential $\mu$. Considering a three-dimensional system, we standardly expand the interaction term in angular harmonics using Legendre polynomials $\wp_l(\cos \theta^{'})$~\cite{Abrikosov-1959,Conti-1999,Vignale-2005,Valentinis-2020}, and we focus on the spin-symmetric channel for density excitations, characterized by the Landau parameters $F_l^{s}$. The result is
\begin{multline}\label{eq:kin_k_pot2}
\left( q v_{\vec{k},\sigma} \cos{\theta}  -\omega \right) \epsilon_{\vec{k}}(\vec{q},\omega)\\ + q v_{\vec{k},\sigma} \cos{\theta} \int \frac{d \Omega^{'}}{4 \pi} \sum_{l=0}^{+\infty} F_l^{s} \wp_l(\cos \theta^{'}) \epsilon_{\vec{k}^{'}} (\vec{q},\omega)  \\ ={\mathscr{I}}_{coll}(\vec{q},\omega)+{\mathscr{I}}_{r}(\vec{q},\omega)-\frac{e}{m v_F^\ast} q v_{\vec{k},\sigma}  \cos{\theta} \left[\vec{k} \cdot \vec{A}(\vec{q},\omega)\right],
\end{multline}
where $\theta$ is the angle between the wave vector $\vec{q}$ and the quasiparticle velocity $v_{\vec{k},\sigma}$, and $d\Omega$ is the solid angle element in three dimensions. 
We further expand the Fermi-surface displacement function $\epsilon_{\vec{k}}(\vec{q},\omega)$ in the spin-symmetric channel in terms of spherical harmonics $\mathscr{Y}_{l}^{m}(\theta,\phi)$~\cite{Valentinis-2020}:
\begin{equation}\label{eq:exp_displ_FS}
 \epsilon_{\vec{k}}(\vec{q},\omega)=\sum_{l=0}^{+\infty} \sum_{m=-l}^{+l} \epsilon_{l,m}^{s} \mathscr{Y}_{l}^{m}(\theta,\phi).
\end{equation}
In the following we consider the first transverse mode with $m=1$, generating shear currents in the first spin-symmetric interaction channel. Therefore we truncate the sums over $l$ in Eqs.\@ (\ref{eq:kin_k_pot2}) and (\ref{eq:exp_displ_FS}) to $l=1$, so that $\sum_{l=0}^{+\infty} F_l^{s} \wp_l(\cos \alpha) \equiv F_0^S+F_1^S \cos{\alpha}$, and $\epsilon_{\vec{k}}(\vec{q},\omega) =\sum_{l=0}^{+\infty}\epsilon_{l,1}^{s}\mathscr{Y}_{l}^{1}(\theta,\phi) \equiv \epsilon^s(\theta) e^{i \phi}$, where $\epsilon^s(\theta)$ collects the $\theta$-dependent part of the displacement. Integration over the angles $\theta'$ and $\phi'$ in Eq.\@ (\ref{eq:kin_k_pot2}) with the displacement $\epsilon^s(\theta) e^{i \phi}$ gives null result for terms involving $F_0^S$~\cite{Valentinis-2020}, and we are left with 
\begin{multline}\label{eq:kin_k_int_coll_A}
\left( \cos{\theta}  -s \right) \epsilon^s(\theta) e^{i \phi} \\ + \cos{\theta}  \int_0^{\pi} \frac{ \sin{\theta'} d\theta'}{4 \pi} F_1^S  \sin \theta \sin{\theta'} \pi e^{i \phi}  \epsilon^s(\theta') \\  =\frac{1}{q v_F^\ast}{\mathscr{I}}_{coll}(\vec{q},\omega)+\frac{1}{q v_F^\ast}{\mathscr{I}}_{r}(\vec{q},\omega)+\frac{e}{m v_F^\ast}  \cos{\theta} \left[\vec{k} \cdot \vec{A}(\vec{q},\omega)\right],
\end{multline}
where 
\begin{equation}\label{eq:s}
s=\frac{\omega}{q v_F^{\ast}},
\end{equation}
and $v_F^{\ast}=v_F/(1+F_1^S/3)$ is the velocity of quasiparticles on the Fermi surface -- i.e.\@ at the Fermi wave vector $\vec{k} \equiv \vec{k}_F$ -- renormalized with respect to $v_F=\hbar k_F/m$ and $k_F$ by quasiparticle interactions. \\
Eq.\@ (\ref{eq:kin_k_int_coll_A}) is the starting point of our analysis: it describes coherent shear vibrations of the Fermi surface with the dispersion relation (\ref{eq:s}) in response to collisions, interactions, momentum-relaxing scattering and applied vector potentials. Notice that Eq.\@ (\ref{eq:kin_k_int_coll_A}) does not depend on $F_0^S$, since shear deformations generate transverse currents with no net density flow.

In the following sections we solve Eq.\@ (\ref{eq:kin_k_int_coll_A}) for the transverse susceptibility in the Galilean continuum by explicitating the form of the collision integral as prescribed by Fermi-liquid theory~\cite{Lea-1973}. Then, we analyze the effect of momentum-relaxing scattering in relaxation-time approximation~\cite{Conti-1999} considering impurities, Umklapp processes and acoustic phonons as independent scattering sources. \\ \\
\noindent
\subsection{Fermi-liquid transverse susceptibility}

Starting from the kinetic equation (\ref{eq:kin_k_int_coll_A}), we can obtain the transverse (current) susceptibility of the Fermi liquid in closed form thanks to the assumption of only two harmonic components of short-range interactions. 
We first neglect both the collision and the momentum-relaxing integrals, i.e.\@ ${\mathscr{I}}_{coll}(\vec{q},\omega)={\mathscr{I}}_{r}(\vec{q},\omega)=0$, and we assume no interactions. In this case, from the expression of the paramagnetic current density one derives the noninteracting transverse paramagnetic susceptibility~\cite{Valentinis-2020}:
\begin{equation}\label{eq:X_T_0}
\chi_T^0 (\vec{q},\omega)= 3\frac{ n m^{\ast}}{m^2}{\cal I}(s),
\end{equation}
where $m^*=m \left(1+ \frac{F_1^S}{3}\right)$ is the effective mass of Landau quasiparticles, and
\begin{equation}\label{eq:vs_int2}
{\cal I}(s) = -\frac{1}{3} + \frac{1}{2} s^2 + \frac{s}{4} \left( 1-s^2 \right) \ln \left(\frac{s+1}{s-1} \right).
\end{equation}
We now allow for momentum-conserving collisions in the Galilean continuum, i.e.\@ with no momentum relaxation. The conservation of particle number, energy and momentum in collisions imposes constraints on the form of the collision integral ${\mathscr{I}}_{coll}(\vec{q},\omega)$~\cite{Abrikosov-1959}. The latter is written is terms of a single collision time $\tau_c=\tau_c(\omega,T)$, which may depend on frequency $\omega$ and temperature $T$, as \cite{Lea-1973, Valentinis-2020}
\begin{multline}\label{eq:coll_int_transv_2}
\frac{1}{q v_F^\ast}{\mathscr{I}}_{coll}(\vec{q},\omega)={\mathscr{I}}_{coll}\left[\epsilon^s(\theta) e^{i \phi} \right] \\ =-\frac{\epsilon^s(\theta)-\left[ \epsilon^s(\theta)\right]_{\rm{av}}-3 \left[\epsilon^s(\theta) \sin \theta \right]_{\rm{av}} \sin \theta}{i \omega \tau_c} s e^{i \phi}
\end{multline}
where the notation $\left[ \cdot \right]_{\rm{av}}= \int_{0}^{\pi} \frac{ (\sin \theta)^2 d \theta}{4} \cdot$ denotes the angular average with respect to $\theta$.
Notice that for purely shear stresses $\left[ \epsilon^s(\theta)\right]_{\rm{av}}=0$~\cite{Lea-1973}. Again the definition of the paramagnetic current density allows one to derive the interacting transverse paramagnetic susceptibility~\cite{Valentinis-2020}. We have
\begin{multline}\label{eq:Xi_T}
\chi_T(\xi)=\frac{X_T^0(\xi)}{1-3\left(\frac{F_1^S}{3}-\beta \right)\frac{m^2}{3 n m^\ast} X_T^0(\xi) +\beta} \\ =\frac{n}{m} \frac{3\left(1+ \frac{F_1^S}{3} \right) {\cal I}(\xi)}{1-3 \left(\frac{F1^s}{3}-\beta\right) {\cal I}(\xi)+\beta}.
\end{multline}
where 
\begin{subequations}\label{eq:Lea_var}
\begin{equation}\label{eq:Lea_var1}
\xi=s \left(1+\frac{i}{\omega \tau_c} \right),
\end{equation}
\begin{equation}\label{eq:Lea_var3}
\beta=\frac{1}{i \omega \tau_c-1}.
\end{equation}
\end{subequations}
To include damping of transverse currents due to momentum relaxation, we have to utilize the full kinetic equation (\ref{eq:kin_k_int_coll_A}) with the integral ${\mathscr{I}}_{r}(\vec{q},\omega)$. A convenient way to relate the momentum-relaxing transverse susceptibility to the result (\ref{eq:Xi_T}) without relaxation is provided by reference~\onlinecite{Conti-1999}: we define the momentum-relaxing integral in single relaxation time approximation as
\begin{equation}\label{eq:relax_int}
\frac{1}{q v_F^\ast}{\mathscr{I}}_{r}(\vec{q},\omega)=-\frac{\epsilon^s(\theta)-\epsilon^r(\theta)}{i \omega \tau_{K}}s e^{i \phi}.
\end{equation}
According to Eq.\@ (\ref{eq:relax_int}), momentum relaxation tends to restore a ``locally relaxed" equilibrium distribution function, characterized by the displacement $\epsilon^r(\theta)$~\cite{Conti-1999}, which would be in equilibrium without relaxation but in the presence of an effective vector potential $\vec{A}_r(\vec{q},\omega)$ modified by relaxation. The scattering processes described by the time $\tau_{K}$ conserve particle number but not current~\cite{Conti-1999}, and may be $\omega$- and $T$-dependent. This way, we obtain the momentum-relaxing transverse susceptibility~\cite{Valentinis-2020}:
\begin{equation}\label{eq:Xi_T_r}
\chi_{T}(\vec{q},\omega)=\frac{n}{m} \frac{\omega}{\omega+i/\tau_K} \frac{\left(1+\tilde{\beta}\right)\left[1+3\cal{I}(\zeta)\right]}{1-3\left(\frac{F_1^S}{3}-\tilde{\beta}\right)\cal{I}(\zeta)+\tilde{\beta}}.
\end{equation}
where
\begin{equation}\label{eq:zeta_var}
\zeta=\xi+s \frac{i}{\omega \tau_{K}}=s\left( 1+\frac{i}{\omega \tau_c}+\frac{i}{\omega \tau_{K}}\right),
\end{equation}
and $\tilde{\beta}=\beta \xi/\zeta$. We recognize a pole at the denominator of Eqs.\@ (\ref{eq:Xi_T_r}) and (\ref{eq:Xi_T}), in the presence or absence of relaxation respectively: this collective mode is transverse sound, studied and observed in the electrically neutral Fermi liquid realized in $^3$He~\cite{Lea-1973,Roach-1976}. The dispersion relation of transverse sound stems from the condition that the susceptibility diverges for a given real frequency $\omega$ and complex momentum $\vec{q}$:
\begin{equation}\label{eq:Pole_XT}
\chi_T(\vec{q},\omega)^{-1}=0
\end{equation}
Eq.\@ (\ref{eq:Pole_XT}) determines the velocity of the transverse collective mode with first Landau parameter $F_1^S$ and collisions encoded by $\tau_{c}=\tau(\omega,T, F_1^S)$. This mode has been labeled $``$transverse sound$"$ in the liquid-helium literature, although physically it amounts to a shear oscillation of the Fermi surface. The numerical solutions of Eqs.\@ (\ref{eq:Xi_T}) and (\ref{eq:Pole_XT}) are shown schematically in Fig.~\ref{fig-omega-q}a. 

For $F_1^S<6$, there is an $F_1^S$-dependent value of $\omega_{th}$ above which Eq.\@ (\ref{eq:Pole_XT})  does not have solutions. Physically, this means that no Fermi-surface collective modes are allowed for $\omega>\omega_{th}$, and the only possible excitations are incoherent electron-hole pairs. Knowing the dependence $\tau_{c}(\omega,T, F_1^S)$, in the Fermi-liquid model -- i.e.\@ Eq. (\ref{eq:tau_qp_fin2}) -- one finds the numerical solution $\omega(T, F_1^S)$ which is displayed with the red solid curves in Fig.~\ref{fig-omega-F}a,b.
For $F_1^S>6$ Eq.\@ (\ref{eq:Pole_XT}) has solutions for all frequencies, but above a certain threshold frequency $\omega_{\rm{th}}$ the solutions are outside the particle-hole continuum and thus free from Landau damping. This threshold is set by solving Eq.\@ (\ref{eq:Pole_XT}) together with the condition
\begin{equation}\label{eq:No_Landau_damp}
{\omega_{\rm{th}}}{\mathop{\rm Re}\nolimits} {q^{ - 1}} =v_F^{\ast}
\end{equation}
The orange dashed curves in Fig.~\ref{fig-omega-F}a,b represent the numerical solution of $\omega_{th}$ as a function of $F_1^S$.  \\ \\
\noindent
\subsection{Dielectric function and polaritons}

The effect of the long range Coulomb interaction is described by the transverse dielectric function, which can be obtained from the paramagnetic transverse susceptibility (\ref{eq:Xi_T_r}) through the Kubo formula. We have
\begin{multline}\label{eq:epsilon_T_FL_r}
\varepsilon_T(\vec{q},\omega)=1 - \frac{e^2}{\varepsilon_0 \omega ^2}{\chi_T}(\vec{q},\omega ) \\ =1-\frac{(\omega_p)^2}{\omega\left(\omega+i/\tau_K\right)} \frac{\left(1+\tilde{\beta}\right)\left[1+3\mathscr{I}(\zeta)\right]}{1-3\left(\frac{F_1^S}{3}-\tilde{\beta}\right)\mathscr{I}(\zeta)+\tilde{\beta}}.
\end{multline}
where $\omega_p=\sqrt{n e^2/(m \varepsilon_0)}$ is the electron plasma frequency.
Notice that the poles of the dielectric function (\ref{eq:epsilon_T_FL_r}) correspond to the transverse sound dispersion relation (\ref{eq:Pole_XT}), i.e.\@ the shear resonance of the Fermi surface. Due to the charge of the electrons, these resonances couple to radiation through the shear stresses exerted by photons. The radiation-matter interaction gives rise to new collective modes, consisting of the combined shear response of photons and Fermi-surface quasiparticles: the polaritons. Mathematically, polaritons satisfy the self-consistent solutions of Maxwell equations inside the Fermi liquid~\cite{Nozieres-1999}:
\begin{equation}\label{eq:coll_modes_charged}
\frac{q^2c^2}{\omega^2}=\varepsilon_T(\vec{q},\omega)
\end{equation}
Analytical solutions of Eqs.\@ (\ref{eq:epsilon_T_FL_r}) and (\ref{eq:coll_modes_charged}) are available in the regimes $q \rightarrow 0$ and $q\rightarrow \infty$: the leading-term solution in the $q \rightarrow 0$ limit reproduces the phenomenology of viscous charged fluids~\cite{FZVM-2014}, while in the $q \rightarrow \infty$ regime we retrieve the phenomenology of anomalous skin effect \cite{Reuter-1948,Sondheimer-2001,Valentinis-2020}. In the regime $\omega \gg v_F^\ast$ Re$q$, the leading-order expansion of Eq.\@ (\ref{eq:coll_modes_charged}) for $\omega /( v_F^\ast q) \rightarrow \infty$ gives \cite{Valentinis-2020}
\begin{equation}\label{eq:epsilon_prop_gen}
{\varepsilon_T}(\vec{q},\omega ) = 1-\frac{(\omega_p)^2}{\omega^2+i \omega/\tau_K +i \omega \tilde{\nu}(\omega) q^2},
\end{equation}
where 
\begin{equation}\label{eq:nu_tilde}
\tilde{\nu}(\omega)=\left(1+\frac{F_1^S}{3}\right)  \frac{(v_F^{*})^2 \tau_c}{5\left(1+\tau_c/\tau_K-i \omega \tau_c\right)}.
\end{equation}
In the limit $\tau_c \ll \tau_{K} $, that is when momentum-conserving collisions are much more frequent than momentum-relaxing scattering, Eq.\@ (\ref{eq:epsilon_prop_gen}) becomes~\cite{Valentinis-2020}
\begin{multline}\label{eq:eps_T_el_VE_rel}
\epsilon_T(\vec{q},\omega)= 1-\frac{(\omega_p)^2}{\omega^2+i \omega/\tau_K+ i \left(\omega+i/\tau_K\right) \nu(\omega) q^2}\\ \approx 1-\frac{(\omega_p)^2}{\omega^2+i \omega/\tau_K+ i \omega \nu(\omega) q^2},
\end{multline}
where we have defined the generalized shear modulus of the isotropic Fermi liquid~\cite{Conti-1999,Giuliani-2005,Bedell-1982}
\begin{equation}\label{eq:nu}
\nu \left( \omega  \right) = \frac{{1 + F_1^S/3}}{5}\frac{(v_F^\ast)^2}{{1/{\tau _c} - i \omega }}.
\end{equation}
Eqs.\@ (\ref{eq:eps_T_el_VE_rel}) and (\ref{eq:nu}) can be derived from the combination of the linearized Navier-Stokes equation for transverse currents and of Maxwell’s equations under the assumption that $\tau_{c}\ll \tau_{K}$. In the propagating shear regime Eqs.\@ (\ref{eq:coll_modes_charged}) and (\ref{eq:epsilon_prop_gen}) give rise to two degenerate polariton branches for each frequency, characterized by the complex wave vectors \begin{widetext}
\begin{equation}\label{eq:VE_modes_rel}
q^2=\frac{\omega^2 }{2 c^2}  +\frac{i \omega}{2 \nu(\omega)} \left\{1  \pm \frac{\sqrt{\tau_{K} \left\{\omega  \left[c^2+ i \nu(\omega) \omega \right]^2-4 i c^2 \nu(\omega) (\omega_p)^2\right\}+i \left[c^2+i \nu(\omega) \omega \right]^2}}{c^2 \sqrt{\omega  \tau_{K}+i}}\right\},
\end{equation}
\end{widetext}
One solution of Eq.\@ (\ref{eq:VE_modes_rel}) is the usual plasmon-polariton which propagates above $\omega_p$, as also found in the standard Drude model, and the other slower-dispersing root is the shear-polariton. The two roots are schematically shown in Fig.~\ref{fig-omega-q}b. The shear-polariton root of Eq.\@ (\ref{eq:VE_modes_rel}) emerges from the electron-hole continuum, thus ceasing to be Landau-damped, for $\omega>\omega_{th}$. We can obtain a quantitative estimation of the latter condition by reverting to Eq.\@ (\ref{eq:No_Landau_damp}) and noticing that, for $F_1^S>6$, the shear-polariton emerges from the continuum in the regime $\omega \tau_c \gg 1$, Re$q\gg$Im$q$. This allows us to find the analytical solution
\begin{equation}\label{eq:shear_propag}
{\omega _{th}} = \frac{{{\omega _p}}}{c}\frac{{3 v_F^\ast}}{{\sqrt {\left( {F_1^S - 6} \right)\left[ {1 - {{\left( {v_F^\ast/c} \right)}^2}} \right]} }}
\end{equation}
Eq.\@ (\ref{eq:shear_propag}) is shown by the blue solid curves in Fig.~\ref{fig-omega-F}c,d. It confirms the early order-of-magnitude estimate by Nozi\`eres and Pines~\cite{Nozieres-1999}, that is $\mathrm{Re}q=\omega/v_F^\ast \approx \omega_p/ c$ at the boundary with the electron-hole continuum. In the region between the blue solid lines and the red dashed lines in Fig.~\ref{fig-omega-F}c,d, the shear-polariton propagates but it cannot be described by the viscous-liquid phenomenology of Eq.\@ (\ref{eq:eps_T_el_VE_rel}). The dispersion of the shear-polariton is robust against momentum relaxation, as it is negligibly affected by the value of $\tau_K$ in the propagating regime. \\ \\ 
\noindent
\subsection{Collision and momentum relaxation rates}

For the electron-electron momentum-conserving collision time $\tau_c=\tau_c(\omega,T)$ we employ the Fermi-liquid result~\cite{Lawrence-1973,vanderMarel-2011,Coleman-2015}
\begin{multline}\label{eq:tau_qp_fin2}
\frac{\hbar}{\tau_c}=\frac{\pi}{12} \left[ \frac{7}{24} (A_1^S)^2+\frac{5}{8}(A_0^S)^2-\frac{5}{12} A_0^S A_1^S \right] \times \\ \frac{(\hbar \omega)^2+( \pi k_B T)^2 }{ E_F^\ast},
\end{multline}
where $A_l^j=F_l^j/\left[1+F_l^j/\left({2l+1}\right)\right]$. Eq.\@ (\ref{eq:tau_qp_fin2}) holds for an interaction comprising the Landau parameters $F_0^S$ and $F_1^S$, consistently with the assumptions made in obtaining the kinetic equation (\ref{eq:kin_k_int_coll_A}). 
The momentum transfer rate, $1/\tau_K$, results from impurity scattering, $1/\tau_i$, electron-phonon scattering, $1/\tau_{e-ph}$, and Umklapp scattering. The Umklapp term is given by a fraction $\alpha_U\in\left(0,1\right)$ of the electron-electron collisions. Typical values in transition-metal compounds are $\alpha_U\approx 0.5$~\cite{Lawrence-1973}. Consequently
\begin{equation}
\frac{1}{{{\tau _K}}} = \frac{{{\alpha_U}}}{{{\tau _c}}} + \frac{1}{{{\tau _{e - ph}}}} + \frac{1}{{{\tau _i}}}
\end{equation}
We calculate the electron-phonon relaxation rate from the many-body self-energy associated with scattering between electrons and an Einstein phonon branch~\cite{Valentinis-2020}. Explicitly,
\begin{subequations}
\begin{equation}\label{eq:tau_eph}
\frac{\hbar}{\tau_{e-ph}(\omega,T)}=\frac{2}{1-\frac{\lambda \hbar \omega_D}{\pi k_B T} \mathscr{Q}\left( \frac{k_B T}{\hbar \omega_D} \right)} \pi \lambda \hbar \omega_D \mathscr{S} \left(\frac{\omega}{\hbar \omega_D}, \frac{k_B T}{\hbar \omega_D}\right),
\end{equation} 
\begin{equation}\label{eq:R_ac}
\mathscr{Q}(t)=\int_0^1 du u^2 \mathrm{Im}\left\{ \psi'\left(\frac{1}{2}+i \frac{u}{2 \pi t} \right)\right\}
\end{equation}
\begin{equation}\label{eq:el-ph_S}
\mathscr{S} \left(x,t\right)=t^3 \int_0^{\frac{1}{t}} du u^2 \left(\frac{1}{e^{u-\frac{x}{t}}+1}+\frac{1}{e^{u+\frac{x}{t}}+1} +\frac{2}{e^u-1}\right). 
\end{equation}
\end{subequations}
where $\lambda$ is the electron-phonon coupling constant, $\hbar \omega_D$ is the Debye energy, and $\psi'(z)$ is the first derivative of the Digamma function.
For the impurity scattering rate, we assume a positive constant $\tau_{i}$ consistently with first Born approximation. \\ \\
\noindent
\subsection{Thin-film optical transmission in propagating-shear regime}

We now turn to the computational method for the optical transmission of a Fermi-liquid thin film of thickness $d$ in the propagating shear regime. The transmission coefficients at the vacuum/material interfaces follow from the condition that the electromagnetic wave in the vacuum has to match the frequency-degenerate polariton modes, Eq.\@ (\ref{eq:VE_modes_rel}), inside the film. To calculate these coefficients, we exploit the analogy between Fermi-liquid electrodynamics for $\omega\gg v_F^\ast$Re$q$ and the electromagnetic response of viscous charged fluids~\cite{FZVM-2014}: appropriate additional constitutive relations at the slab boundaries can be formulated in terms of the slip length $\lambda$, in accordance with Ref.~\onlinecite{FZVM-2014}, which closes the system of equations for the reflection and transmission coefficients of each polariton mode at the slab boundaries. A total of six linear equations stem from the continuity of the electric field and of its derivative, and the constitutive relation from the linearized Navier-Stokes equation. The solution of such linear equations gives the transmission coefficient of the film $t_{\rm film}=t_{\rm film}(\omega)$ relative to the transmission in vacuum $t_v$:
\begin{multline}\label{eq:t_PS}
t_{ \rm film}\left( \omega  \right)/{t_v} = {e^{ - i\omega d/c}}\left[ {t_ + }{e^{i{q_ + }d}} + {\theta _ + }{e^{ - i{q_ + }d}} + {t_ - }{e^{i{q_ - }d}} \right. \\ \left. + {\theta _ - }{e^{ - i{q_ - }d}} \right],
\end{multline}
where the indices refer to the $ \pm $  branches of Eq.\@ (\ref{eq:VE_modes_rel}), $t_v$ and the amplitudes $t_\pm$, $\theta_\pm$ are determined numerically through Eq.\@ (9) of Ref.~\onlinecite{FZVM-2014}.\\
We microscopically determine the slip length following Ref.~\onlinecite{Kiselev-2019}:
\begin{subequations}\label{eq:lambda_s}
\begin{equation}\label{eq:lambda_s_s}
\lambda^{s}=\frac{1}{h^2 (h')^2 (k_F)^4} \frac{45 \pi^2}{(k_F)^4 \hbar} n m \nu(0) =\frac{3}{h^2 (h')^2 (k_F)^4} v_F^{*} \tau_c ,
\end{equation}
\begin{equation}\label{eq:lambda_s_d}
\lambda^{d}=\frac{8 \pi^2}{(k_F)^4 \hbar} n m \nu(0)=\frac{8}{15} v_F^{*} \tau_c,
\end{equation}
\end{subequations}
where $h$ and $h^{\prime}$ are amplitude and correlation length of the interface roughness, $k_F$ is the Fermi wave vector, and $ \lambda^s$ ($ \lambda^d$) refer to specular (diffuse) scattering. In Fig.~\ref{fig-t-omega}c-f we compare the results for $t(\omega)=t_{ \rm film}\left( \omega  \right)/{t_v}$ from the propagating-shear model, Eq.\@ (\ref{eq:t_PS}), with the standard Drude model (Ohmic conductor) for thin-film transmission.


\begin{thebibliography}{44}%
\makeatletter
\providecommand \@ifxundefined [1]{%
 \@ifx{#1\undefined}
}%
\providecommand \@ifnum [1]{%
 \ifnum #1\expandafter \@firstoftwo
 \else \expandafter \@secondoftwo
 \fi
}%
\providecommand \@ifx [1]{%
 \ifx #1\expandafter \@firstoftwo
 \else \expandafter \@secondoftwo
 \fi
}%
\providecommand \natexlab [1]{#1}%
\providecommand \enquote  [1]{``#1''}%
\providecommand \bibnamefont  [1]{#1}%
\providecommand \bibfnamefont [1]{#1}%
\providecommand \citenamefont [1]{#1}%
\providecommand \href@noop [0]{\@secondoftwo}%
\providecommand \href [0]{\begingroup \@sanitize@url \@href}%
\providecommand \@href[1]{\@@startlink{#1}\@@href}%
\providecommand \@@href[1]{\endgroup#1\@@endlink}%
\providecommand \@sanitize@url [0]{\catcode `\\12\catcode `\$12\catcode
  `\&12\catcode `\#12\catcode `\^12\catcode `\_12\catcode `\%12\relax}%
\providecommand \@@startlink[1]{}%
\providecommand \@@endlink[0]{}%
\providecommand \url  [0]{\begingroup\@sanitize@url \@url }%
\providecommand \@url [1]{\endgroup\@href {#1}{\urlprefix }}%
\providecommand \urlprefix  [0]{URL }%
\providecommand \Eprint [0]{\href }%
\providecommand \doibase [0]{http://dx.doi.org/}%
\providecommand \selectlanguage [0]{\@gobble}%
\providecommand \bibinfo  [0]{\@secondoftwo}%
\providecommand \bibfield  [0]{\@secondoftwo}%
\providecommand \translation [1]{[#1]}%
\providecommand \BibitemOpen [0]{}%
\providecommand \bibitemStop [0]{}%
\providecommand \bibitemNoStop [0]{.\EOS\space}%
\providecommand \EOS [0]{\spacefactor3000\relax}%
\providecommand \BibitemShut  [1]{\csname bibitem#1\endcsname}%
\let\auto@bib@innerbib\@empty
\bibitem [{\citenamefont {Nozi\`{e}res}\ and\ \citenamefont
  {Pines}(1999)}]{Nozieres-1999}%
  \BibitemOpen
  \bibfield  {author} {\bibinfo {author} {\bibfnamefont {P.}~\bibnamefont
  {Nozi\`{e}res}}\ and\ \bibinfo {author} {\bibfnamefont {D.}~\bibnamefont
  {Pines}},\ }\href@noop {} {\emph {\bibinfo {title} {Theory Of Quantum
  Liquids}}},\ Advanced Books Classics Series\ (\bibinfo  {publisher} {Westview
  Press},\ \bibinfo {address} {Nashville},\ \bibinfo {year} {1999})\BibitemShut
  {NoStop}%
\bibitem [{\citenamefont {Abel}\ \emph {et~al.}(1966)\citenamefont {Abel},
  \citenamefont {Anderson},\ and\ \citenamefont {Wheatley}}]{Abel-1966}%
  \BibitemOpen
  \bibfield  {author} {\bibinfo {author} {\bibfnamefont {W.~R.}\ \bibnamefont
  {Abel}}, \bibinfo {author} {\bibfnamefont {A.~C.}\ \bibnamefont {Anderson}},
  \ and\ \bibinfo {author} {\bibfnamefont {J.~C.}\ \bibnamefont {Wheatley}},\
  }\href {\doibase 10.1103/PhysRevLett.17.74} {\bibfield  {journal} {\bibinfo
  {journal} {Phys. Rev. Lett.}\ }\textbf {\bibinfo {volume} {17}},\ \bibinfo
  {pages} {74} (\bibinfo {year} {1966})}\BibitemShut {NoStop}%
\bibitem [{\citenamefont {Abrikosov}\ and\ \citenamefont
  {Khalatnikov}(1959)}]{Abrikosov-1959}%
  \BibitemOpen
  \bibfield  {author} {\bibinfo {author} {\bibfnamefont {A.~A.}\ \bibnamefont
  {Abrikosov}}\ and\ \bibinfo {author} {\bibfnamefont {I.~M.}\ \bibnamefont
  {Khalatnikov}},\ }\href {http://stacks.iop.org/0034-4885/22/i=1/a=310}
  {\bibfield  {journal} {\bibinfo  {journal} {Rep. Prog. Phys.}\ }\textbf
  {\bibinfo {volume} {22}},\ \bibinfo {pages} {329} (\bibinfo {year}
  {1959})}\BibitemShut {NoStop}%
\bibitem [{\citenamefont {Lea}\ \emph {et~al.}(1973)\citenamefont {Lea},
  \citenamefont {Birks}, \citenamefont {Lee},\ and\ \citenamefont
  {Dobbs}}]{Lea-1973}%
  \BibitemOpen
  \bibfield  {author} {\bibinfo {author} {\bibfnamefont {M.~J.}\ \bibnamefont
  {Lea}}, \bibinfo {author} {\bibfnamefont {A.~R.}\ \bibnamefont {Birks}},
  \bibinfo {author} {\bibfnamefont {P.~M.}\ \bibnamefont {Lee}}, \ and\
  \bibinfo {author} {\bibfnamefont {E.~R.}\ \bibnamefont {Dobbs}},\ }\href
  {http://stacks.iop.org/0022-3719/6/i=11/a=004} {\bibfield  {journal}
  {\bibinfo  {journal} {J. Phys. C: Solid State}\ }\textbf {\bibinfo {volume}
  {6}},\ \bibinfo {pages} {L226} (\bibinfo {year} {1973})}\BibitemShut
  {NoStop}%
\bibitem [{\citenamefont {Beekman}\ \emph
  {et~al.}(2017{\natexlab{a}})\citenamefont {Beekman}, \citenamefont
  {Nissinen}, \citenamefont {Wu}, \citenamefont {Liu}, \citenamefont {Slager},
  \citenamefont {Nussinov}, \citenamefont {Cvetkovic},\ and\ \citenamefont
  {Zaanen}}]{Beekman-2017a}%
  \BibitemOpen
  \bibfield  {author} {\bibinfo {author} {\bibfnamefont {A.~J.}\ \bibnamefont
  {Beekman}}, \bibinfo {author} {\bibfnamefont {J.}~\bibnamefont {Nissinen}},
  \bibinfo {author} {\bibfnamefont {K.}~\bibnamefont {Wu}}, \bibinfo {author}
  {\bibfnamefont {K.}~\bibnamefont {Liu}}, \bibinfo {author} {\bibfnamefont
  {R.-J.}\ \bibnamefont {Slager}}, \bibinfo {author} {\bibfnamefont
  {Z.}~\bibnamefont {Nussinov}}, \bibinfo {author} {\bibfnamefont
  {V.}~\bibnamefont {Cvetkovic}}, \ and\ \bibinfo {author} {\bibfnamefont
  {J.}~\bibnamefont {Zaanen}},\ }\href
  {http://www.sciencedirect.com/science/article/pii/S0370157317300583}
  {\bibfield  {journal} {\bibinfo  {journal} {Phys. Rep.}\ }\textbf {\bibinfo
  {volume} {683}},\ \bibinfo {pages} {1 } (\bibinfo {year}
  {2017}{\natexlab{a}})}\BibitemShut {NoStop}%
\bibitem [{\citenamefont {Beekman}\ \emph
  {et~al.}(2017{\natexlab{b}})\citenamefont {Beekman}, \citenamefont
  {Nissinen}, \citenamefont {Wu},\ and\ \citenamefont
  {Zaanen}}]{Beekman-2017b}%
  \BibitemOpen
  \bibfield  {author} {\bibinfo {author} {\bibfnamefont {A.~J.}\ \bibnamefont
  {Beekman}}, \bibinfo {author} {\bibfnamefont {J.}~\bibnamefont {Nissinen}},
  \bibinfo {author} {\bibfnamefont {K.}~\bibnamefont {Wu}}, \ and\ \bibinfo
  {author} {\bibfnamefont {J.}~\bibnamefont {Zaanen}},\ }\href {\doibase
  10.1103/PhysRevB.96.165115} {\bibfield  {journal} {\bibinfo  {journal} {Phys.
  Rev. B}\ }\textbf {\bibinfo {volume} {96}},\ \bibinfo {pages} {165115}
  (\bibinfo {year} {2017}{\natexlab{b}})}\BibitemShut {NoStop}%
\bibitem [{\citenamefont {Conti}\ and\ \citenamefont
  {Vignale}(1999)}]{Conti-1999}%
  \BibitemOpen
  \bibfield  {author} {\bibinfo {author} {\bibfnamefont {S.}~\bibnamefont
  {Conti}}\ and\ \bibinfo {author} {\bibfnamefont {G.}~\bibnamefont
  {Vignale}},\ }\href {\doibase 10.1103/PhysRevB.60.7966} {\bibfield  {journal}
  {\bibinfo  {journal} {Phys. Rev. B}\ }\textbf {\bibinfo {volume} {60}},\
  \bibinfo {pages} {7966} (\bibinfo {year} {1999})}\BibitemShut {NoStop}%
\bibitem [{\citenamefont {Roach}\ and\ \citenamefont
  {Ketterson}(1976)}]{Roach-1976}%
  \BibitemOpen
  \bibfield  {author} {\bibinfo {author} {\bibfnamefont {P.~R.}\ \bibnamefont
  {Roach}}\ and\ \bibinfo {author} {\bibfnamefont {J.~B.}\ \bibnamefont
  {Ketterson}},\ }\href {\doibase 10.1103/PhysRevLett.36.736} {\bibfield
  {journal} {\bibinfo  {journal} {Phys. Rev. Lett.}\ }\textbf {\bibinfo
  {volume} {36}},\ \bibinfo {pages} {736} (\bibinfo {year} {1976})}\BibitemShut
  {NoStop}%
\bibitem [{\citenamefont {Stricker}\ \emph {et~al.}(2014)\citenamefont
  {Stricker}, \citenamefont {Mravlje}, \citenamefont {Berthod}, \citenamefont
  {Fittipaldi}, \citenamefont {Vecchione}, \citenamefont {Georges},\ and\
  \citenamefont {van~der Marel}}]{Stricker-2014}%
  \BibitemOpen
  \bibfield  {author} {\bibinfo {author} {\bibfnamefont {D.}~\bibnamefont
  {Stricker}}, \bibinfo {author} {\bibfnamefont {J.}~\bibnamefont {Mravlje}},
  \bibinfo {author} {\bibfnamefont {C.}~\bibnamefont {Berthod}}, \bibinfo
  {author} {\bibfnamefont {R.}~\bibnamefont {Fittipaldi}}, \bibinfo {author}
  {\bibfnamefont {A.}~\bibnamefont {Vecchione}}, \bibinfo {author}
  {\bibfnamefont {A.}~\bibnamefont {Georges}}, \ and\ \bibinfo {author}
  {\bibfnamefont {D.}~\bibnamefont {van~der Marel}},\ }\href {\doibase
  10.1103/PhysRevLett.113.087404} {\bibfield  {journal} {\bibinfo  {journal}
  {Phys. Rev. Lett.}\ }\textbf {\bibinfo {volume} {113}},\ \bibinfo {pages}
  {087404} (\bibinfo {year} {2014})}\BibitemShut {NoStop}%
\bibitem [{\citenamefont {Khoo}\ and\ \citenamefont
  {Villadiego}(2019)}]{Khoo-2019}%
  \BibitemOpen
  \bibfield  {author} {\bibinfo {author} {\bibfnamefont {J.~Y.}\ \bibnamefont
  {Khoo}}\ and\ \bibinfo {author} {\bibfnamefont {I.~S.}\ \bibnamefont
  {Villadiego}},\ }\href {\doibase 10.1103/PhysRevB.99.075434} {\bibfield
  {journal} {\bibinfo  {journal} {Phys. Rev. B}\ }\textbf {\bibinfo {volume}
  {99}},\ \bibinfo {pages} {075434} (\bibinfo {year} {2019})}\BibitemShut
  {NoStop}%
\bibitem [{\citenamefont {Khoo}\ \emph {et~al.}(2020)\citenamefont {Khoo},
  \citenamefont {Chang}, \citenamefont {Pientka},\ and\ \citenamefont
  {Sodemann}}]{Khoo-2020}%
  \BibitemOpen
  \bibfield  {author} {\bibinfo {author} {\bibfnamefont {J.~Y.}\ \bibnamefont
  {Khoo}}, \bibinfo {author} {\bibfnamefont {P.-Y.}\ \bibnamefont {Chang}},
  \bibinfo {author} {\bibfnamefont {F.}~\bibnamefont {Pientka}}, \ and\
  \bibinfo {author} {\bibfnamefont {I.}~\bibnamefont {Sodemann}},\ }\href
  {\doibase 10.1103/PhysRevB.102.085437} {\bibfield  {journal} {\bibinfo
  {journal} {Phys. Rev. B}\ }\textbf {\bibinfo {volume} {102}},\ \bibinfo
  {pages} {085437} (\bibinfo {year} {2020})}\BibitemShut {NoStop}%
\bibitem [{Note1()}]{Note1}%
  \BibitemOpen
  \bibinfo {note} {For all calculations in this work we employed the
  momentum-conserving collision time $\tau _c$ (a single-particle quantity) in
  accordance with Eq.\spacefactor \@m {} (24) in sec.\spacefactor \@m {} \ref
  {methods}. This is different from the sound-attenuation or optical scattering
  rates (two-particle quantities), which for a Fermi liquid are $ \propto
  \mathord {\mathchar "077E h}E_F/[ (\mathord {\mathchar "077E h}\omega
  )^2+(2\pi k_B T)^2]$ \cite {Berthod-2013}.}\BibitemShut {Stop}%
\bibitem [{\citenamefont {Valentinis}(2020)}]{Valentinis-2020}%
  \BibitemOpen
  \bibfield  {author} {\bibinfo {author} {\bibfnamefont {D.}~\bibnamefont
  {Valentinis}},\ }\href {https://arxiv.org/abs/2003.06619} {\bibfield
  {journal} {\bibinfo  {journal} {arXiv:2003.06619}\ } (\bibinfo {year}
  {2020})}\BibitemShut {NoStop}%
\bibitem [{\citenamefont {van~der Marel}\ \emph {et~al.}(2011)\citenamefont
  {van~der Marel}, \citenamefont {van Mechelen},\ and\ \citenamefont
  {Mazin}}]{vanderMarel-2011}%
  \BibitemOpen
  \bibfield  {author} {\bibinfo {author} {\bibfnamefont {D.}~\bibnamefont
  {van~der Marel}}, \bibinfo {author} {\bibfnamefont {J.~L.~M.}\ \bibnamefont
  {van Mechelen}}, \ and\ \bibinfo {author} {\bibfnamefont {I.~I.}\
  \bibnamefont {Mazin}},\ }\href {\doibase 10.1103/PhysRevB.84.205111}
  {\bibfield  {journal} {\bibinfo  {journal} {Phys. Rev. B}\ }\textbf {\bibinfo
  {volume} {84}},\ \bibinfo {pages} {205111} (\bibinfo {year}
  {2011})}\BibitemShut {NoStop}%
\bibitem [{\citenamefont {Forcella}\ \emph
  {et~al.}(2014{\natexlab{a}})\citenamefont {Forcella}, \citenamefont {Zaanen},
  \citenamefont {Valentinis},\ and\ \citenamefont {van~der Marel}}]{FZVM-2014}%
  \BibitemOpen
  \bibfield  {author} {\bibinfo {author} {\bibfnamefont {D.}~\bibnamefont
  {Forcella}}, \bibinfo {author} {\bibfnamefont {J.}~\bibnamefont {Zaanen}},
  \bibinfo {author} {\bibfnamefont {D.}~\bibnamefont {Valentinis}}, \ and\
  \bibinfo {author} {\bibfnamefont {D.}~\bibnamefont {van~der Marel}},\ }\href
  {\doibase 10.1103/PhysRevB.90.035143} {\bibfield  {journal} {\bibinfo
  {journal} {Phys. Rev. B}\ }\textbf {\bibinfo {volume} {90}},\ \bibinfo
  {pages} {035143} (\bibinfo {year} {2014}{\natexlab{a}})}\BibitemShut
  {NoStop}%
\bibitem [{\citenamefont {Crossno}\ \emph {et~al.}(2016)\citenamefont
  {Crossno}, \citenamefont {Shi}, \citenamefont {Wang}, \citenamefont {Liu},
  \citenamefont {Harzheim}, \citenamefont {Lucas}, \citenamefont {Sachdev},
  \citenamefont {Kim}, \citenamefont {Taniguchi}, \citenamefont {Watanabe}
  \emph {et~al.}}]{Crossno-2016}%
  \BibitemOpen
  \bibfield  {author} {\bibinfo {author} {\bibfnamefont {J.}~\bibnamefont
  {Crossno}}, \bibinfo {author} {\bibfnamefont {J.~K.}\ \bibnamefont {Shi}},
  \bibinfo {author} {\bibfnamefont {K.}~\bibnamefont {Wang}}, \bibinfo {author}
  {\bibfnamefont {X.}~\bibnamefont {Liu}}, \bibinfo {author} {\bibfnamefont
  {A.}~\bibnamefont {Harzheim}}, \bibinfo {author} {\bibfnamefont
  {A.}~\bibnamefont {Lucas}}, \bibinfo {author} {\bibfnamefont
  {S.}~\bibnamefont {Sachdev}}, \bibinfo {author} {\bibfnamefont
  {P.}~\bibnamefont {Kim}}, \bibinfo {author} {\bibfnamefont {T.}~\bibnamefont
  {Taniguchi}}, \bibinfo {author} {\bibfnamefont {K.}~\bibnamefont {Watanabe}},
   \emph {et~al.},\ }\href
  {https://science.sciencemag.org/content/351/6277/1058} {\bibfield  {journal}
  {\bibinfo  {journal} {Science}\ }\textbf {\bibinfo {volume} {351}},\ \bibinfo
  {pages} {1058} (\bibinfo {year} {2016})}\BibitemShut {NoStop}%
\bibitem [{\citenamefont {Bandurin}\ \emph {et~al.}(2016)\citenamefont
  {Bandurin}, \citenamefont {Torre}, \citenamefont {Krishna~Kumar},
  \citenamefont {Ben~Shalom}, \citenamefont {Tomadin}, \citenamefont
  {Principi}, \citenamefont {Auton}, \citenamefont {Khestanova}, \citenamefont
  {Novoselov}, \citenamefont {Grigorieva}, \citenamefont {Ponomarenko},
  \citenamefont {Geim},\ and\ \citenamefont {Polini}}]{Bandurin-2016}%
  \BibitemOpen
  \bibfield  {author} {\bibinfo {author} {\bibfnamefont {D.~A.}\ \bibnamefont
  {Bandurin}}, \bibinfo {author} {\bibfnamefont {I.}~\bibnamefont {Torre}},
  \bibinfo {author} {\bibfnamefont {R.}~\bibnamefont {Krishna~Kumar}}, \bibinfo
  {author} {\bibfnamefont {M.}~\bibnamefont {Ben~Shalom}}, \bibinfo {author}
  {\bibfnamefont {A.}~\bibnamefont {Tomadin}}, \bibinfo {author} {\bibfnamefont
  {A.}~\bibnamefont {Principi}}, \bibinfo {author} {\bibfnamefont
  {G.}~\bibnamefont {Auton}}, \bibinfo {author} {\bibfnamefont
  {E.}~\bibnamefont {Khestanova}}, \bibinfo {author} {\bibfnamefont {K.~S.}\
  \bibnamefont {Novoselov}}, \bibinfo {author} {\bibfnamefont {I.~V.}\
  \bibnamefont {Grigorieva}}, \bibinfo {author} {\bibfnamefont {L.~A.}\
  \bibnamefont {Ponomarenko}}, \bibinfo {author} {\bibfnamefont {A.~K.}\
  \bibnamefont {Geim}}, \ and\ \bibinfo {author} {\bibfnamefont
  {M.}~\bibnamefont {Polini}},\ }\href
  {https://science.sciencemag.org/content/351/6277/1055} {\bibfield  {journal}
  {\bibinfo  {journal} {Science}\ }\textbf {\bibinfo {volume} {351}},\ \bibinfo
  {pages} {1055} (\bibinfo {year} {2016})}\BibitemShut {NoStop}%
\bibitem [{\citenamefont {Kumar}\ \emph {et~al.}(2017)\citenamefont {Kumar},
  \citenamefont {Bandurin}, \citenamefont {Pellegrino}, \citenamefont {Cao},
  \citenamefont {Principi}, \citenamefont {Guo}, \citenamefont {Auton},
  \citenamefont {Shalom}, \citenamefont {Ponomarenko}, \citenamefont
  {Falkovich}, \citenamefont {Watanabe}, \citenamefont {Taniguchi},
  \citenamefont {Grigorieva}, \citenamefont {Levitov}, \citenamefont {Polini},\
  and\ \citenamefont {Geim}}]{Kumar-2017}%
  \BibitemOpen
  \bibfield  {author} {\bibinfo {author} {\bibfnamefont {K.~R.}\ \bibnamefont
  {Kumar}}, \bibinfo {author} {\bibfnamefont {D.}~\bibnamefont {Bandurin}},
  \bibinfo {author} {\bibfnamefont {F.}~\bibnamefont {Pellegrino}}, \bibinfo
  {author} {\bibfnamefont {Y.}~\bibnamefont {Cao}}, \bibinfo {author}
  {\bibfnamefont {A.}~\bibnamefont {Principi}}, \bibinfo {author}
  {\bibfnamefont {H.}~\bibnamefont {Guo}}, \bibinfo {author} {\bibfnamefont
  {G.}~\bibnamefont {Auton}}, \bibinfo {author} {\bibfnamefont {M.~B.}\
  \bibnamefont {Shalom}}, \bibinfo {author} {\bibfnamefont {L.~A.}\
  \bibnamefont {Ponomarenko}}, \bibinfo {author} {\bibfnamefont
  {G.}~\bibnamefont {Falkovich}}, \bibinfo {author} {\bibfnamefont
  {K.}~\bibnamefont {Watanabe}}, \bibinfo {author} {\bibfnamefont
  {T.}~\bibnamefont {Taniguchi}}, \bibinfo {author} {\bibfnamefont {I.~V.}\
  \bibnamefont {Grigorieva}}, \bibinfo {author} {\bibfnamefont
  {L.}~\bibnamefont {Levitov}}, \bibinfo {author} {\bibfnamefont
  {M.}~\bibnamefont {Polini}}, \ and\ \bibinfo {author} {\bibfnamefont
  {A.}~\bibnamefont {Geim}},\ }\href
  {https://www.nature.com/articles/nphys4240} {\bibfield  {journal} {\bibinfo
  {journal} {Nat. Phys.}\ } (\bibinfo {year} {2017})}\BibitemShut {NoStop}%
\bibitem [{\citenamefont {Gooth}\ \emph {et~al.}(2018)\citenamefont {Gooth},
  \citenamefont {Menges}, \citenamefont {Kumar}, \citenamefont {S{\"u}$\beta$},
  \citenamefont {Shekhar}, \citenamefont {Sun}, \citenamefont {Drechsler},
  \citenamefont {Zierold}, \citenamefont {Felser},\ and\ \citenamefont
  {Gotsmann}}]{Gooth-2018}%
  \BibitemOpen
  \bibfield  {author} {\bibinfo {author} {\bibfnamefont {J.}~\bibnamefont
  {Gooth}}, \bibinfo {author} {\bibfnamefont {F.}~\bibnamefont {Menges}},
  \bibinfo {author} {\bibfnamefont {N.}~\bibnamefont {Kumar}}, \bibinfo
  {author} {\bibfnamefont {V.}~\bibnamefont {S{\"u}$\beta$}}, \bibinfo {author}
  {\bibfnamefont {C.}~\bibnamefont {Shekhar}}, \bibinfo {author} {\bibfnamefont
  {Y.}~\bibnamefont {Sun}}, \bibinfo {author} {\bibfnamefont {U.}~\bibnamefont
  {Drechsler}}, \bibinfo {author} {\bibfnamefont {R.}~\bibnamefont {Zierold}},
  \bibinfo {author} {\bibfnamefont {C.}~\bibnamefont {Felser}}, \ and\ \bibinfo
  {author} {\bibfnamefont {B.}~\bibnamefont {Gotsmann}},\ }\href
  {https://www.nature.com/articles/s41467-018-06688-y} {\bibfield  {journal}
  {\bibinfo  {journal} {Nat. Comm.}\ }\textbf {\bibinfo {volume} {9}} (\bibinfo
  {year} {2018})}\BibitemShut {NoStop}%
\bibitem [{\citenamefont {Sulpizio}\ \emph {et~al.}(2019)\citenamefont
  {Sulpizio}, \citenamefont {Ella}, \citenamefont {Rozen}, \citenamefont
  {Birkbeck}, \citenamefont {Perello}, \citenamefont {Dutta}, \citenamefont
  {Ben-Shalom}, \citenamefont {Taniguchi}, \citenamefont {Watanabe},
  \citenamefont {Holder}, \citenamefont {Queiroz}, \citenamefont {Principi},
  \citenamefont {Stern}, \citenamefont {Scaffidi}, \citenamefont {Geim},\ and\
  \citenamefont {Ilani}}]{Sulpizio-2019}%
  \BibitemOpen
  \bibfield  {author} {\bibinfo {author} {\bibfnamefont {J.~A.}\ \bibnamefont
  {Sulpizio}}, \bibinfo {author} {\bibfnamefont {L.}~\bibnamefont {Ella}},
  \bibinfo {author} {\bibfnamefont {A.}~\bibnamefont {Rozen}}, \bibinfo
  {author} {\bibfnamefont {J.}~\bibnamefont {Birkbeck}}, \bibinfo {author}
  {\bibfnamefont {D.~J.}\ \bibnamefont {Perello}}, \bibinfo {author}
  {\bibfnamefont {D.}~\bibnamefont {Dutta}}, \bibinfo {author} {\bibfnamefont
  {M.}~\bibnamefont {Ben-Shalom}}, \bibinfo {author} {\bibfnamefont
  {T.}~\bibnamefont {Taniguchi}}, \bibinfo {author} {\bibfnamefont
  {K.}~\bibnamefont {Watanabe}}, \bibinfo {author} {\bibfnamefont
  {T.}~\bibnamefont {Holder}}, \bibinfo {author} {\bibfnamefont
  {R.}~\bibnamefont {Queiroz}}, \bibinfo {author} {\bibfnamefont
  {A.}~\bibnamefont {Principi}}, \bibinfo {author} {\bibfnamefont
  {A.}~\bibnamefont {Stern}}, \bibinfo {author} {\bibfnamefont
  {T.}~\bibnamefont {Scaffidi}}, \bibinfo {author} {\bibfnamefont {A.~K.}\
  \bibnamefont {Geim}}, \ and\ \bibinfo {author} {\bibfnamefont
  {S.}~\bibnamefont {Ilani}},\ }\href
  {https://www.nature.com/articles/s41586-019-1788-9} {\bibfield  {journal}
  {\bibinfo  {journal} {Nature}\ }\textbf {\bibinfo {volume} {576}},\ \bibinfo
  {pages} {75} (\bibinfo {year} {2019})}\BibitemShut {NoStop}%
\bibitem [{\citenamefont {Ella}\ \emph {et~al.}(2019)\citenamefont {Ella},
  \citenamefont {Rozen}, \citenamefont {Birkbeck}, \citenamefont {Ben-Shalom},
  \citenamefont {Perello}, \citenamefont {Zultak}, \citenamefont {Taniguchi},
  \citenamefont {Watanabe}, \citenamefont {Geim}, \citenamefont {Ilani},\ and\
  \citenamefont {Sulpizio}}]{Ella-2019}%
  \BibitemOpen
  \bibfield  {author} {\bibinfo {author} {\bibfnamefont {L.}~\bibnamefont
  {Ella}}, \bibinfo {author} {\bibfnamefont {A.}~\bibnamefont {Rozen}},
  \bibinfo {author} {\bibfnamefont {J.}~\bibnamefont {Birkbeck}}, \bibinfo
  {author} {\bibfnamefont {M.}~\bibnamefont {Ben-Shalom}}, \bibinfo {author}
  {\bibfnamefont {D.}~\bibnamefont {Perello}}, \bibinfo {author} {\bibfnamefont
  {J.}~\bibnamefont {Zultak}}, \bibinfo {author} {\bibfnamefont
  {T.}~\bibnamefont {Taniguchi}}, \bibinfo {author} {\bibfnamefont
  {K.}~\bibnamefont {Watanabe}}, \bibinfo {author} {\bibfnamefont {A.~K.}\
  \bibnamefont {Geim}}, \bibinfo {author} {\bibfnamefont {S.}~\bibnamefont
  {Ilani}}, \ and\ \bibinfo {author} {\bibfnamefont {J.~A.}\ \bibnamefont
  {Sulpizio}},\ }\href {https://www.nature.com/articles/s41565-019-0398-x}
  {\bibfield  {journal} {\bibinfo  {journal} {Nat. Nanotechnol.}\ ,\ \bibinfo
  {pages} {1}} (\bibinfo {year} {2019})}\BibitemShut {NoStop}%
\bibitem [{\citenamefont {Silin}(1958{\natexlab{a}})}]{Silin-1958a}%
  \BibitemOpen
  \bibfield  {author} {\bibinfo {author} {\bibfnamefont {V.~P.}\ \bibnamefont
  {Silin}},\ }\href {http://www.jetp.ac.ru/cgi-bin/dn/e_006_02_0387.pdf}
  {\bibfield  {journal} {\bibinfo  {journal} {Sov. Phys. JETP-USSR}\ }\textbf
  {\bibinfo {volume} {6}},\ \bibinfo {pages} {387} (\bibinfo {year}
  {1958}{\natexlab{a}})}\BibitemShut {NoStop}%
\bibitem [{\citenamefont {Silin}(1958{\natexlab{b}})}]{Silin-1958b}%
  \BibitemOpen
  \bibfield  {author} {\bibinfo {author} {\bibfnamefont {V.}~\bibnamefont
  {Silin}},\ }\href {http://www.jetp.ac.ru/cgi-bin/dn/e_006_05_0985.pdf}
  {\bibfield  {journal} {\bibinfo  {journal} {Sov. Phys. JETP-USSR}\ }\textbf
  {\bibinfo {volume} {6}},\ \bibinfo {pages} {985} (\bibinfo {year}
  {1958}{\natexlab{b}})}\BibitemShut {NoStop}%
\bibitem [{\citenamefont {Forcella}\ \emph
  {et~al.}(2014{\natexlab{b}})\citenamefont {Forcella}, \citenamefont
  {Mezzalira},\ and\ \citenamefont {Musso}}]{Forcella-2014}%
  \BibitemOpen
  \bibfield  {author} {\bibinfo {author} {\bibfnamefont {D.}~\bibnamefont
  {Forcella}}, \bibinfo {author} {\bibfnamefont {A.}~\bibnamefont {Mezzalira}},
  \ and\ \bibinfo {author} {\bibfnamefont {D.}~\bibnamefont {Musso}},\ }\href
  {\doibase 10.1007/JHEP11(2014)153} {\bibfield  {journal} {\bibinfo  {journal}
  {J. High Energy Phys.}\ }\textbf {\bibinfo {volume} {2014}},\ \bibinfo {pages}
  {1} (\bibinfo {year} {2014}{\natexlab{b}})}\BibitemShut {NoStop}%
\bibitem [{\citenamefont {Baggioli}\ \emph {et~al.}(2020)\citenamefont
  {Baggioli}, \citenamefont {Gran},\ and\ \citenamefont
  {Torns{\"o}}}]{Baggioli-2020}%
  \BibitemOpen
  \bibfield  {author} {\bibinfo {author} {\bibfnamefont {M.}~\bibnamefont
  {Baggioli}}, \bibinfo {author} {\bibfnamefont {U.}~\bibnamefont {Gran}}, \
  and\ \bibinfo {author} {\bibfnamefont {M.}~\bibnamefont {Torns{\"o}}},\
  }\href {https://link.springer.com/article/10.1007/JHEP04(2020)106} {\bibfield
   {journal} {\bibinfo  {journal} {J. High Energy Phys.}\ }\textbf {\bibinfo
  {volume} {2020}},\ \bibinfo {pages} {1} (\bibinfo {year} {2020})}\BibitemShut
  {NoStop}%
\bibitem [{\citenamefont {Reuter}\ and\ \citenamefont
  {Sondheimer}(1948)}]{Reuter-1948}%
  \BibitemOpen
  \bibfield  {author} {\bibinfo {author} {\bibfnamefont {G.~E.~H.}\
  \bibnamefont {Reuter}}\ and\ \bibinfo {author} {\bibfnamefont {E.~H.}\
  \bibnamefont {Sondheimer}},\ }\href {\doibase 10.1098/rspa.1948.0123}
  {\bibfield  {journal} {\bibinfo  {journal} {Proc. R. Soc. A}\ }\textbf
  {\bibinfo {volume} {195}},\ \bibinfo {pages} {336} (\bibinfo {year}
  {1948})}\BibitemShut {NoStop}%
\bibitem [{\citenamefont {Gurzhi}(1968)}]{Gurzhi-1968}%
  \BibitemOpen
  \bibfield  {author} {\bibinfo {author} {\bibfnamefont {R.~N.}\ \bibnamefont
  {Gurzhi}},\ }\href {http://stacks.iop.org/0038-5670/11/i=2/a=R07} {\bibfield
  {journal} {\bibinfo  {journal} {Sov. Phys. Uspekhi}\ }\textbf {\bibinfo
  {volume} {11}},\ \bibinfo {pages} {255} (\bibinfo {year} {1968})}\BibitemShut
  {NoStop}%
\bibitem [{\citenamefont {Valentinis}\ \emph {et~al.}(2020)\citenamefont
  {Valentinis}, \citenamefont {Zaanen},\ and\ \citenamefont {van~der
  Marel}}]{Impedance-unpubl-2020}%
  \BibitemOpen
  \bibfield  {author} {\bibinfo {author} {\bibfnamefont {D.}~\bibnamefont
  {Valentinis}}, \bibinfo {author} {\bibfnamefont {J.}~\bibnamefont {Zaanen}},
  \ and\ \bibinfo {author} {\bibfnamefont {D.}~\bibnamefont {van~der Marel}},\
  }\href@noop {} {} (\bibinfo {year} {2020}),\ \bibinfo {note}
  {unpublished}\BibitemShut {NoStop}%
\bibitem [{\citenamefont {Coleman}(2010)}]{Coleman-2015}%
  \BibitemOpen
  \bibfield  {author} {\bibinfo {author} {\bibfnamefont {P.}~\bibnamefont
  {Coleman}},\ }\href@noop {} {\emph {\bibinfo {title} {Introduction to many
  body physics}}}\ (\bibinfo  {publisher} {Cambridge University Press},\
  \bibinfo {address} {Cambridge},\ \bibinfo {year} {2010})\BibitemShut
  {NoStop}%
\bibitem [{\citenamefont {Kotliar}\ \emph {et~al.}(2006)\citenamefont
  {Kotliar}, \citenamefont {Savrasov}, \citenamefont {Haule}, \citenamefont
  {Oudovenko}, \citenamefont {Parcollet},\ and\ \citenamefont
  {Marianetti}}]{Kotliar-2006}%
  \BibitemOpen
  \bibfield  {author} {\bibinfo {author} {\bibfnamefont {G.}~\bibnamefont
  {Kotliar}}, \bibinfo {author} {\bibfnamefont {S.~Y.}\ \bibnamefont
  {Savrasov}}, \bibinfo {author} {\bibfnamefont {K.}~\bibnamefont {Haule}},
  \bibinfo {author} {\bibfnamefont {V.~S.}\ \bibnamefont {Oudovenko}}, \bibinfo
  {author} {\bibfnamefont {O.}~\bibnamefont {Parcollet}}, \ and\ \bibinfo
  {author} {\bibfnamefont {C.~A.}\ \bibnamefont {Marianetti}},\ }\href
  {\doibase 10.1103/RevModPhys.78.865} {\bibfield  {journal} {\bibinfo
  {journal} {Rev. Mod. Phys.}\ }\textbf {\bibinfo {volume} {78}},\ \bibinfo
  {pages} {865} (\bibinfo {year} {2006})}\BibitemShut {NoStop}%
\bibitem [{\citenamefont {Bradlyn}\ \emph {et~al.}(2012)\citenamefont
  {Bradlyn}, \citenamefont {Goldstein},\ and\ \citenamefont
  {Read}}]{Bradlyn-2012}%
  \BibitemOpen
  \bibfield  {author} {\bibinfo {author} {\bibfnamefont {B.}~\bibnamefont
  {Bradlyn}}, \bibinfo {author} {\bibfnamefont {M.}~\bibnamefont {Goldstein}},
  \ and\ \bibinfo {author} {\bibfnamefont {N.}~\bibnamefont {Read}},\ }\href
  {\doibase 10.1103/PhysRevB.86.245309} {\bibfield  {journal} {\bibinfo
  {journal} {Phys. Rev. B}\ }\textbf {\bibinfo {volume} {86}},\ \bibinfo
  {pages} {245309} (\bibinfo {year} {2012})}\BibitemShut {NoStop}%
\bibitem [{\citenamefont {Cook}\ and\ \citenamefont {Lucas}(2019)}]{Cook-2019}%
  \BibitemOpen
  \bibfield  {author} {\bibinfo {author} {\bibfnamefont {C.~Q.}\ \bibnamefont
  {Cook}}\ and\ \bibinfo {author} {\bibfnamefont {A.}~\bibnamefont {Lucas}},\
  }\href {\doibase 10.1103/PhysRevB.99.235148} {\bibfield  {journal} {\bibinfo
  {journal} {Phys. Rev. B}\ }\textbf {\bibinfo {volume} {99}},\ \bibinfo
  {pages} {235148} (\bibinfo {year} {2019})}\BibitemShut {NoStop}%
\bibitem [{\citenamefont {van~der Eb}\ \emph {et~al.}(2001)\citenamefont
  {van~der Eb}, \citenamefont {Kuzmenko},\ and\ \citenamefont {van~der
  Marel}}]{vanderEb-2001}%
  \BibitemOpen
  \bibfield  {author} {\bibinfo {author} {\bibfnamefont {J.~W.}\ \bibnamefont
  {van~der Eb}}, \bibinfo {author} {\bibfnamefont {A.~B.}\ \bibnamefont
  {Kuzmenko}}, \ and\ \bibinfo {author} {\bibfnamefont {D.}~\bibnamefont
  {van~der Marel}},\ }\href {\doibase 10.1103/PhysRevLett.86.3407} {\bibfield
  {journal} {\bibinfo  {journal} {Phys. Rev. Lett.}\ }\textbf {\bibinfo
  {volume} {86}},\ \bibinfo {pages} {3407} (\bibinfo {year}
  {2001})}\BibitemShut {NoStop}%
\bibitem [{\citenamefont {Haule}\ \emph {et~al.}(2005)\citenamefont {Haule},
  \citenamefont {Oudovenko}, \citenamefont {Savrasov},\ and\ \citenamefont
  {Kotliar}}]{Haule-2005}%
  \BibitemOpen
  \bibfield  {author} {\bibinfo {author} {\bibfnamefont {K.}~\bibnamefont
  {Haule}}, \bibinfo {author} {\bibfnamefont {V.}~\bibnamefont {Oudovenko}},
  \bibinfo {author} {\bibfnamefont {S.~Y.}\ \bibnamefont {Savrasov}}, \ and\
  \bibinfo {author} {\bibfnamefont {G.}~\bibnamefont {Kotliar}},\ }\href
  {\doibase 10.1103/PhysRevLett.94.036401} {\bibfield  {journal} {\bibinfo
  {journal} {Phys. Rev. Lett.}\ }\textbf {\bibinfo {volume} {94}},\ \bibinfo
  {pages} {036401} (\bibinfo {year} {2005})}\BibitemShut {NoStop}%
\bibitem [{\citenamefont {Mena}\ \emph {et~al.}(2003)\citenamefont {Mena},
  \citenamefont {van~der Marel}, \citenamefont {Damascelli}, \citenamefont
  {F\"ath}, \citenamefont {Menovsky},\ and\ \citenamefont
  {Mydosh}}]{Mena-2003}%
  \BibitemOpen
  \bibfield  {author} {\bibinfo {author} {\bibfnamefont {F.~P.}\ \bibnamefont
  {Mena}}, \bibinfo {author} {\bibfnamefont {D.}~\bibnamefont {van~der Marel}},
  \bibinfo {author} {\bibfnamefont {A.}~\bibnamefont {Damascelli}}, \bibinfo
  {author} {\bibfnamefont {M.}~\bibnamefont {F\"ath}}, \bibinfo {author}
  {\bibfnamefont {A.~A.}\ \bibnamefont {Menovsky}}, \ and\ \bibinfo {author}
  {\bibfnamefont {J.~A.}\ \bibnamefont {Mydosh}},\ }\href {\doibase
  10.1103/PhysRevB.67.241101} {\bibfield  {journal} {\bibinfo  {journal} {Phys.
  Rev. B}\ }\textbf {\bibinfo {volume} {67}},\ \bibinfo {pages} {241101}
  (\bibinfo {year} {2003})}\BibitemShut {NoStop}%
\bibitem [{\citenamefont {{Scheffler, M. and Dressel, M. and Jourdan, M. and
  Adrian, H.}}(2005)}]{Scheffler-2005}%
  \BibitemOpen
  \bibfield  {author} {\bibinfo {author} {\bibnamefont {{Scheffler, M. and
  Dressel, M. and Jourdan, M. and Adrian, H.}}},\ }\href
  {https://www.nature.com/articles/nature04232} {\bibfield  {journal} {\bibinfo
   {journal} {Nature}\ }\textbf {\bibinfo {volume} {438}},\ \bibinfo {pages}
  {1135} (\bibinfo {year} {2005})}\BibitemShut {NoStop}%
\bibitem [{\citenamefont {Prochaska}\ \emph {et~al.}(2020)\citenamefont
  {Prochaska}, \citenamefont {Li}, \citenamefont {MacFarland}, \citenamefont
  {Andrews}, \citenamefont {Bonta}, \citenamefont {Bianco}, \citenamefont
  {Yazdi}, \citenamefont {Schrenk}, \citenamefont {Detz}, \citenamefont
  {Limbeck}, \citenamefont {Si}, \citenamefont {Ringe}, \citenamefont
  {Strasser3}, \citenamefont {Kono},\ and\ \citenamefont
  {Paschen}}]{Prochaska-2020}%
  \BibitemOpen
  \bibfield  {author} {\bibinfo {author} {\bibfnamefont {L.}~\bibnamefont
  {Prochaska}}, \bibinfo {author} {\bibfnamefont {X.}~\bibnamefont {Li}},
  \bibinfo {author} {\bibfnamefont {D.~C.}\ \bibnamefont {MacFarland}},
  \bibinfo {author} {\bibfnamefont {A.~M.}\ \bibnamefont {Andrews}}, \bibinfo
  {author} {\bibfnamefont {M.}~\bibnamefont {Bonta}}, \bibinfo {author}
  {\bibfnamefont {E.~F.}\ \bibnamefont {Bianco}}, \bibinfo {author}
  {\bibfnamefont {S.}~\bibnamefont {Yazdi}}, \bibinfo {author} {\bibfnamefont
  {W.}~\bibnamefont {Schrenk}}, \bibinfo {author} {\bibfnamefont
  {H.}~\bibnamefont {Detz}}, \bibinfo {author} {\bibfnamefont {A.}~\bibnamefont
  {Limbeck}}, \bibinfo {author} {\bibfnamefont {Q.}~\bibnamefont {Si}},
  \bibinfo {author} {\bibfnamefont {E.}~\bibnamefont {Ringe}}, \bibinfo
  {author} {\bibfnamefont {G.}~\bibnamefont {Strasser3}}, \bibinfo {author}
  {\bibfnamefont {J.}~\bibnamefont {Kono}}, \ and\ \bibinfo {author}
  {\bibfnamefont {S.}~\bibnamefont {Paschen}},\ }\href
  {https://science.sciencemag.org/content/367/6475/285} {\bibfield  {journal}
  {\bibinfo  {journal} {Science}\ }\textbf {\bibinfo {volume} {367}},\ \bibinfo
  {pages} {285} (\bibinfo {year} {2020})}\BibitemShut {NoStop}%
\bibitem [{\citenamefont {Giuliani}\ and\ \citenamefont
  {Vignale}(2005{\natexlab{a}})}]{Vignale-2005}%
  \BibitemOpen
  \bibfield  {author} {\bibinfo {author} {\bibfnamefont {G.}~\bibnamefont
  {Giuliani}}\ and\ \bibinfo {author} {\bibfnamefont {G.}~\bibnamefont
  {Vignale}},\ }\href@noop {} {\emph {\bibinfo {title} {Quantum Theory of the
  Electron Liquid}}}\ (\bibinfo  {publisher} {Cambridge University Press},\
  \bibinfo {address} {Cambridge},\ \bibinfo {year} {2005})\BibitemShut
  {NoStop}%
\bibitem [{\citenamefont {Sondheimer}(2001)}]{Sondheimer-2001}%
  \BibitemOpen
  \bibfield  {author} {\bibinfo {author} {\bibfnamefont {E.~H.}\ \bibnamefont
  {Sondheimer}},\ }\href {\doibase 10.1080/00018730152707225} {\bibfield
  {journal} {\bibinfo  {journal} {Ad. Phys.}\ }\textbf {\bibinfo {volume}
  {50}},\ \bibinfo {pages} {499} (\bibinfo {year} {2001})}\BibitemShut
  {NoStop}%
\bibitem [{\citenamefont {Giuliani}\ and\ \citenamefont
  {Vignale}(2005{\natexlab{b}})}]{Giuliani-2005}%
  \BibitemOpen
  \bibfield  {author} {\bibinfo {author} {\bibfnamefont {G.}~\bibnamefont
  {Giuliani}}\ and\ \bibinfo {author} {\bibfnamefont {G.}~\bibnamefont
  {Vignale}},\ }\href@noop {} {\emph {\bibinfo {title} {Quantum Theory of the
  Electron Liquid}}}\ (\bibinfo  {publisher} {Cambridge University Press},\
  \bibinfo {address} {Cambridge},\ \bibinfo {year} {2005})\BibitemShut
  {NoStop}%
\bibitem [{\citenamefont {Bedell}\ and\ \citenamefont
  {Pethick}(1982)}]{Bedell-1982}%
  \BibitemOpen
  \bibfield  {author} {\bibinfo {author} {\bibfnamefont {K.}~\bibnamefont
  {Bedell}}\ and\ \bibinfo {author} {\bibfnamefont {C.~J.}\ \bibnamefont
  {Pethick}},\ }\href {\doibase 10.1007/BF00681588} {\bibfield  {journal}
  {\bibinfo  {journal} {J. Low Temp. Phys.}\ }\textbf {\bibinfo {volume}
  {49}},\ \bibinfo {pages} {213} (\bibinfo {year} {1982})}\BibitemShut
  {NoStop}%
\bibitem [{\citenamefont {Lawrence}\ and\ \citenamefont
  {Wilkins}(1973)}]{Lawrence-1973}%
  \BibitemOpen
  \bibfield  {author} {\bibinfo {author} {\bibfnamefont {W.~E.}\ \bibnamefont
  {Lawrence}}\ and\ \bibinfo {author} {\bibfnamefont {J.~W.}\ \bibnamefont
  {Wilkins}},\ }\href {\doibase 10.1103/PhysRevB.7.2317} {\bibfield  {journal}
  {\bibinfo  {journal} {Phys. Rev. B}\ }\textbf {\bibinfo {volume} {7}},\
  \bibinfo {pages} {2317} (\bibinfo {year} {1973})}\BibitemShut {NoStop}%
\bibitem [{\citenamefont {Kiselev}\ and\ \citenamefont
  {Schmalian}(2019)}]{Kiselev-2019}%
  \BibitemOpen
  \bibfield  {author} {\bibinfo {author} {\bibfnamefont {E.~I.}\ \bibnamefont
  {Kiselev}}\ and\ \bibinfo {author} {\bibfnamefont {J.}~\bibnamefont
  {Schmalian}},\ }\href {\doibase 10.1103/PhysRevB.99.035430} {\bibfield
  {journal} {\bibinfo  {journal} {Phys. Rev. B}\ }\textbf {\bibinfo {volume}
  {99}},\ \bibinfo {pages} {035430} (\bibinfo {year} {2019})}\BibitemShut
  {NoStop}%
\bibitem [{\citenamefont {Berthod}\ \emph {et~al.}(2013)\citenamefont
  {Berthod}, \citenamefont {Mravlje}, \citenamefont {Deng}, \citenamefont
  {\ifmmode~\check{Z}\else \v{Z}\fi{}itko}, \citenamefont {van~der Marel},\
  and\ \citenamefont {Georges}}]{Berthod-2013}%
  \BibitemOpen
  \bibfield  {author} {\bibinfo {author} {\bibfnamefont {C.}~\bibnamefont
  {Berthod}}, \bibinfo {author} {\bibfnamefont {J.}~\bibnamefont {Mravlje}},
  \bibinfo {author} {\bibfnamefont {X.}~\bibnamefont {Deng}}, \bibinfo {author}
  {\bibfnamefont {R.}~\bibnamefont {\ifmmode~\check{Z}\else \v{Z}\fi{}itko}},
  \bibinfo {author} {\bibfnamefont {D.}~\bibnamefont {van~der Marel}}, \ and\
  \bibinfo {author} {\bibfnamefont {A.}~\bibnamefont {Georges}},\ }\href
  {\doibase 10.1103/PhysRevB.87.115109} {\bibfield  {journal} {\bibinfo
  {journal} {Phys. Rev. B}\ }\textbf {\bibinfo {volume} {87}},\ \bibinfo
  {pages} {115109} (\bibinfo {year} {2013})}\BibitemShut {NoStop}%
\end{thebibliography}
\end{document}